\documentclass[prd, twocolumn, lengthcheck, superscriptaddress, showpacs, letterpaper, nofootinbib]{revtex4-1}
\usepackage{hyperref}
\usepackage{cleveref}
\usepackage{times}
\usepackage{amsmath}
\usepackage{graphicx}
\DeclareGraphicsExtensions{.pdf,.eps,.png,.jpg}
\usepackage{acronym}
\usepackage{color}
\usepackage[caption=false]{subfig}

\newcommand\given{\,\,|\,}

\newcommand\auxvec{x}
\newcommand\Vdata{V_\mathrm{d}}

\newcommand\MLArank{r_{\textsc{MLA}}}
\newcommand\ANNrank{r_{\textsc{ANN}}}
\newcommand\SVMrank{r_{\textsc{SVM}}}
\newcommand\RFrank{r_{\textsc{RF}}}
\newcommand\OVLrank{r_{\textsc{OVL}}}
\newcommand\vecrank{\vec{r}}

\newcommand\MLAmax{\ac{MLA}${}_{\mathrm{max}}$}

\newcommand\Ft{\ensuremath{F^\ast}}
\newcommand\Pt{\ensuremath{P_0^\ast}}

\newcommand\auxvect{\auxvec^{\ast}}
\newcommand\pt{\ensuremath{p(\auxvect)}}
\newcommand\qt{\ensuremath{q(\auxvect)}}

\newcommand\pVg{\ensuremath{p(V_1)}}
\newcommand\pVc{\ensuremath{p(V_0)}}

\newcommand\diff{\, \mathrm{d}}

\def\thercsid{\relax}
\def\rcsid#1{\def\next##1#1{\def\thercsid{##1}}\next}
\rcsid$Id: auxmvc.tex,v 1.86 2013/03/27 19:44:35 reed.essick Exp $

\newcommand\tsinghuaDCST{\affiliation{Department of Computer Science
    and Technology, Tsinghua University, Beijing 100084, P. R. China}}

\newcommand\tsinghuaRIIT{\affiliation{Research Institute of
    Information Technology, Tsinghua National Laboratory for
    Information Science and Technology, Tsinghua University, Beijing
    100084, P. R. China }}

\begin{document}

\title{Application of machine learning algorithms to the study of noise artifacts in gravitational-wave data}

\author{Rahul Biswas}
\affiliation{University of Texas-Brownsville, Brownsville, Texas 78520, USA} 

\author{Lindy Blackburn}
\affiliation{NASA Goddard Space Flight Center, Greenbelt, MD 20771, USA}

\author{Junwei Cao}
\tsinghuaRIIT

\author{Reed Essick}
\affiliation{LIGO - Massachusetts Institute of Technology, Cambridge, MA 02139, USA}

\author{Kari Alison Hodge}
\affiliation{LIGO - California Institute of Technology, Pasadena, CA 91125, USA}

\author{Erotokritos Katsavounidis}
\affiliation{LIGO - Massachusetts Institute of Technology, Cambridge, MA 02139, USA}

\author{Kyungmin Kim}
\affiliation{Hanyang University, Seoul 133-791, Korea}
\affiliation{National Institute for Mathematical Sciences, Daejeon 305-811, Korea}

\author{Young-Min Kim}
\affiliation{Pusan National University, Busan 609-735, Korea}
\affiliation{National Institute for Mathematical Sciences, Daejeon 305-811, Korea}

\author{Eric-Olivier Le~Bigot}
\tsinghuaRIIT

\author{Chang-Hwan Lee}
\affiliation{Pusan National University, Busan 609-735, Korea}

\author{John J. Oh}
\affiliation{National Institute for Mathematical Sciences, Daejeon 305-811, Korea}

\author{Sang Hoon Oh}
\affiliation{National Institute for Mathematical Sciences, Daejeon 305-811, Korea}

\author{Edwin J. Son}
\affiliation{National Institute for Mathematical Sciences, Daejeon 305-811, Korea}

\author{Ruslan Vaulin}
\affiliation{LIGO - Massachusetts Institute of Technology, Cambridge, MA 02139, USA}

\author{Xiaoge Wang}
\tsinghuaDCST

\author{Tao Ye}
\tsinghuaDCST

%\date{\thercsid}

\begin{abstract}

The sensitivity of searches for astrophysical transients in data from the Laser
Interferometer Gravitational-wave Observatory (LIGO) is generally limited by
the presence of transient, non-Gaussian noise artifacts, which occur at a
high-enough rate such that accidental coincidence across multiple detectors is
non-negligible. Furthermore, non-Gaussian noise artifacts typically dominate over the background contributed from stationary
noise. These ``glitches'' can easily be confused for transient gravitational-wave
signals, and their robust identification and removal will help any search for
astrophysical gravitational-waves. We apply Machine Learning Algorithms (MLAs)
to the problem, using data from auxiliary channels within the LIGO detectors
that monitor degrees of freedom unaffected by astrophysical signals.
Terrestrial noise sources may manifest characteristic disturbances in these
auxiliary channels, inducing non-trivial correlations with glitches in the
gravitational-wave data.  The number of auxiliary-channel parameters describing these
disturbances may also be extremely large; an area where MLAs are particularly
well-suited.  We demonstrate the feasibility and applicability of three very
different MLAs: Artificial Neural Networks, Support Vector Machines, and
Random Forests. These classifiers identify and remove a substantial fraction of
the glitches present in two very different data sets: four weeks of LIGO's
fourth science run and one week of LIGO's sixth science run. We observe that
all three algorithms agree on which events are glitches to within 10\% for the 
sixth science run data, and support this by showing that
the different optimization criteria used by each classifier generate the same
decision surface, based on a likelihood-ratio statistic.  Furthermore, we find
that all classifiers obtain similar limiting performance, suggesting that most
of the useful information currently contained in the auxiliary channel parameters we
extract is already being used.  Future performance gains are thus likely to
involve additional sources of information, rather than improvements in the
MLAs themselves.

\end{abstract}

\maketitle
%acronyms
%\acrodef{SNR}{signal-to-noise ratio}
\acrodef{LIGO}{Laser Interferometer Gravitational-wave Observatory}
\acrodef{MLA}{Machine learning algorithms}
\acrodef{ROC}{Receiver Operating Characteristic}
\acrodef{ANN}{Artificial Neural Network}
\acrodef{SVM}{Support Vector Machines}
\acrodef{RF}{Random Forest}
\acrodef{OVL}{Ordered Veto List}
\acrodef{MLP}{multi-layered perceptron}
\acrodef{GA}{genetic algorithm}
\acrodef{RBF}{Radial Basis function}
\acrodef{cdf}{cumulative density functions}
\acrodef{pdf}{probability density function}
\acrodef{FANN}{Fast Artificial Neural Network}

%%%%%%%%%%%%%%%%%%%%%%%%%%%%%%%%%%%%%%%%%%%%%%%%%%%%%%%%%%%%%%%%%%%%%%%%
\section{Introduction}
\label{sec:intro}
%%%%%%%%%%%%%%%%%%%%%%%%%%%%%%%%%%%%%%%%%%%%%%%%%%%%%%%%%%%%%%%%%%%%%%%%

The \ac{LIGO} is a two-site network of ground-based detectors designed for the direct detection and measurement of gravitational-wave signals from astrophysical sources~\cite{Abbott:2007kv, 0264-9381-27-8-084006}. The \ac{LIGO} detectors, in their initial configuration~\cite{Abbott:2007kv}, have operated since 2001 and conducted several scientific runs, collecting data with incrementally increased sensitivity in each run~\cite{Abbott:2007kv,PSD:S5,Collaboration:2012wu}. Although no gravitational waves were detected, these runs tested and refined key technologies, as well as provided a large amount of data characterizing the detectors. The next generation of detectors, referred to as the advanced \ac{LIGO} detectors, are currently under construction and are expected to be operational by 2015~\cite{Abbott:2007kv, 0264-9381-27-8-084006}. Major upgrades to lasers, optics, and seismic isolation/sensing will provide roughly a factor of ten improvement to sensitivity, which corresponds to a factor of 1000 in the observable volume of space and the number of detectable sources. Based on our current knowledge of potential astrophysical sources, the advanced \ac{LIGO} detectors are expected to make routine gravitational-wave detections (see for example~\cite{ratesdoc}) and will open the era of gravitational-wave astronomy. 

The \ac{LIGO} detector noise may be characterized by an approximately stationary component of
colored Gaussian noise, with the addition of short duration non-Gaussian noise
artifacts called ``glitches'' (other noise sources such as non-stationary lines
and broadband non-stationarity do not always fit neatly into this framework).
The stationary noise in the instrument is dominated by low-frequency seismic
noise coupling to mirror motion, thermal noise in the mirrors and suspensions,
60~Hz power lines and harmonics, and shot noise.  Sources of transient noise
can include temporary seismic, acoustic, or magnetic disturbances, power
transients, scattered light, dust crossing the beam, instabilities in the
interferometer, channel saturations, and other complicated and often non-linear
effects. To monitor these disturbances and keep the instrument in a stable
operating condition through active feedback, each detector records hundreds of
auxiliary channels along with the gravitational-wave channel.  These auxiliary
channels keep track of important non-gravitational-wave degrees of freedom in
the interferometer, as well as information about the local environment. They
are critical to understanding the state of the instrument at any particular
time.

One of \ac{LIGO}'s main scientific goals is the detection of transient
gravitational-wave signals, which can come from the coalescence of a compact binary or
core-collapse supernova, among other astrophysical sources~\cite{Cutler:2001}.
The presence of glitches is problematic for searches targeting these signals
because glitches can be easily confused with transient gravitational-wave signals.
The primary method to distinguish a real gravitational-wave transient from
instrumental artifact is to check that a signal appears in two or more
geographically distant detectors. While this coincidence requirement is
extremely effective, a high rate of glitches means that accidental coincidence
of noise transients across multiple detectors still dominates the search
background, resulting in weaker upper limits and making the confident detection
of real signals challenging. The problem is most severe in searches for
transients with poorly-modeled or little identifying waveform structure, such
as generic gravitational-wave bursts or intermediate-mass binary black-hole
coalescence which spend only a short amount of time (few cycles) in the LIGO
sensitive band during merger.

While the precise noise characteristics in the advanced detectors will be
different from those of initial \ac{LIGO}, glitch sources for future data will
exist and the detection problem for short duration signals will persist. Thus,
it is critical to develop data analysis methods for the robust identification
of glitches in \ac{LIGO} data. For previous investigations of instrumental
artifacts, their impact on gravitational-wave searches and the methods for
their identification,
see~\cite{DiCredico2005,Ajith2007,Blackburn2008,Tomoki2010,Christensen:2010,Slutsky:2010short,Smith:2011an,Virgodetchar2012}.
We use the \ac{OVL} algorithm as a benchmark for our investigations. \ac{OVL}
has been used in recent \ac{LIGO} science runs as one of the primary glitch
detection algorithms. In particular, an earlier version of \ac{OVL} described
in \cite{blackburnthesis} was used during \ac{LIGO}'s fifth science run
\cite{Abbott:2009zi}. \ac{OVL} attempts to measure the degree of likelihood
that a gravitational wave candidate can be associated with a transient
instrumental disturbance found in one of the many auxiliary channels.

Glitches are induced by the detector's environment, noise in the detector
subsystems, or a combination of thereof. These sources should appear in the
auxiliary channels as well. Because legitimate gravitational waves may couple
to channels besides the nominal gravitational wave channel, we use a subset of
auxiliary channels shown to be insensitive to gravitational waves. This subset
is generated through hardware injections at the detectors~\cite{Smith:2011an}.
The hardware injections involve driving the test masses through magnetic
coupling with an expected gravitational wave signal and searching for evidence
of that signal in auxiliary channels. If the signal does not systematically
appear in an auxiliary channel, that channel is deemed ``safe'' and we include
it in our analysis. By analyzing information from these auxiliary channels, one
may be able to distinguish glitches from genuine gravitational-wave signals and
ideally establish their cause. The main difficulty in such an analysis is
processing the information from hundreds of channels which may manifest
non-trivial correlations between themselves when they respond to an
instrumental disturbance. Given the high dimensionality and the absence of
reliable statistical models for noise and coupling between auxiliary channels,
traditional computational methods are not well-suited to this problem. On the
other hand, \ac{MLA}s have been used to solve problems like this since the
1970's in other fields such as computer science, biology, and finance. 

This paper presents the use of \ac{MLA}s for the purpose of glitch identification in gravitational-wave detectors. The main goal of the paper is to establish the feasibility of applying \ac{MLA}s in the context of the \ac{LIGO} detectors. We consider three well-known algorithms: the \ac{ANN}, the \ac{SVM}, and the \ac{RF}. We explore their properties and test their performance by analyzing data from past scientific runs. Based on  these tests, we discuss the prospects for using \ac{MLA}s for glitch identification in the advanced \ac{LIGO} detectors.

This paper is organized as follows. In Section~\ref{Dataprep}, we describe the process for reducing raw time-series data and preparing feature vectors for the \ac{MLA} classifiers. This is followed by a general formulation of the glitch detection problem in Section~\ref{detection_problem}. Then, in Section~\ref{pipeline}, we briefly describe the classifiers'  algorithms. Training and testing of the classifiers is discussed in Section~\ref{tuning}. Finally, we evaluate and compare the classifiers' performances using the standard \ac{ROC} curves in Section~\ref{comparison} and investigate various ways of combining classifiers in Section~\ref{combining}. In Appendix~\ref{appendix:fom}, we explore several optimization criteria used by the classifiers and verify their theoretical consistency.  

%%%%%%%%%%%%%%%%%%%%%%%%%%%%%%%%%%%%%%%%%%%%%%%%%%%%
\section{Data Preparation}
\label{Dataprep}
%%%%%%%%%%%%%%%%%%%%%%%%%%%%%%%%%%%%%%%%%%%%%%%%%%%%

We use data taken by the 4km-arm detector at Hanford, WA (H1) during \ac{LIGO}'s fourth science run (S4: 24 February -- 24 March 2005), and data taken by the 4km-arm detector at Livingston, LA (L1) during one week (28 May -- 4 June 2010) of \ac{LIGO}'s sixth science run (S6: 7 July 2009 -- 20 October 2010). Hereafter we refer to these data sets as the S4 and the S6 data. 

In the time between the fourth and the sixth science runs, the detectors
underwent major commissioning and improvements to their sensitivity. Thus,
while the H1 and L1 detectors are identical by design, the data taken by H1
during S4 and the data taken by L1 during S6 are quite different. These data
sets represent evolutionary changes in both the detector noise
power-spectral-density and the non-Gaussian transient artifacts.
Differences in the detectors' environments due to their distant geographical
locations add another degree of freedom. Processing data from detectors
separated in time and location allows us to determine how adaptable and robust
these analysis algorithms are. This is important when extrapolating their
performance to advanced detectors.         

Classification, or the separation of input data into various categories, is one of MLAs main uses; thus, they are often referred to as classifiers. We have two categories of data: glitches (Class 1) and ``clean'' data (Class 0). If one was to perform a search for gravitational-wave transient signals, the first category, glitches, would generally be identified as candidate transient events and considered false alarms. The second category, ``clean'' data, contain only Gaussian detector noise. A true gravitational-wave signal, when it arrives at the detector, is superposed on the Gaussian detector noise. If signal's amplitude is high enough, it also would be identified by the search algorithm as a candidate transient event. Since it is a genuine gravitational-wave transient, it would constitute an actual detection, as opposed to glitches which act as noise. Hereafter we refer to such candidate gravitational-wave transient events, either genuine gravitational-wave transients or glitches, as transient events or, simply, as  events. 

We characterize a transient event in either class by information from the detector's auxiliary channels. Importantly, we record the same information for both classes of events. Each channel records a time-series measuring some non-gravitational-wave degree of freedom, either in the detector or its environment. We first reduce the time-series to a set of non-Gaussian transients using the Kleine Welle analysis algorithm~\cite{ref:omegagrams}, which finds clusters of excess signal power in the dyadic wavelet domain. The detected transients are ranked by their statistical significance, defined as the negative logarithm of the probability that a random cluster of wavelet coefficients subject to Gaussian noise would contain the same or greater signal power. The \ac{MLA} classifiers use the properties of auxiliary channel transients coincident in time with the gravitational-wave channel event to classify the gravitational-wave event.

Given an event in the gravitational-wave channel at time $t$, we build a feature vector $\auxvec$ out of the nearby auxiliary channel transients.
Each channel contributes five features:
\begin{itemize}
	\item $\rho$: The significance of the single loudest transient in that auxiliary channel within 
              $\pm$100 ms of $t$.
	\item $\Delta t$: The difference between $t$ and the central time corresponding to the auxiliary channel transient.
	\item $d$: The duration of the auxiliary channel transient.
	\item $f$: The central frequency of the auxiliary channel transient.
	\item $n$: The number of wavelet coefficients clustered to form the auxiliary channel 
              transient (a measure of time-frequency volume).
\end{itemize}
We require all auxiliary transients to have a significance of at least 15, and if no such auxiliary transient is found within 100 ms of $t$, the five fields for that channel are set to zero. The significance threshold of 15 and 100 ms window are tunable parameters. The 100 ms window covers most transient coupling timescales identified by previous studies~\cite{Smith:2011an}. 
However, there is no guarantee that this window is an optimal choice, or that it should be the same for all auxiliary channels. 
In total, we analyze 250 (162) auxiliary channels from S6 (S4) data, resulting in 1250 (810) dimensions for the auxiliary feature vector, $\auxvec$. In addition, we record certain bookkeeping information about the original gravitational-wave channel event, the state of nearby non-Gaussian transients in the gravitational-wave channel, and other information about data quality.  
These values are stripped before classifier training and evaluation so that we train the classifiers on only information contained in the auxiliary features.

The set of ``glitch'' (Class 1) samples, $\{\auxvec\}_1$, is generated by running Kleine Welle over the gravitational-wave channel from one of the \ac{LIGO} detectors. 
This set of non-Gaussian transients from the gravitational-wave channel can, in principle, contain true gravitational-waves. However, without requiring coincident transients in another detector, they are overwhelmingly dominated by noise artifacts. Even for the most sensitive data set (S6), the expected rate of detectable gravitational-wave transients from known astrophysical sources is extremely low (${\sim}10^{-9}$ Hz~\cite{ratesdoc}) with respect to the rate of single-detector noise transients (${\sim}0.1$ Hz). For the advanced LIGO detectors, it may be appropriate to remove coincident gravitational-wave candidates from the glitch training samples to avoid contamination from detectable gravitational-wave events.
In both classifiers' training and performance evaluation, we treat all Kleine Welle transients from the gravitational-wave channel as artifacts. In total, we identify 2832 (16,204) noise transients above a nominal significance threshold of 35 from the Livingston L1 (Hanford H1) detector during one week of the S6 (four weeks of the S4) science run. The central time from each event is used to trigger feature vector generation, so that $\{\auxvec\}_1$ is a set of 2832 (16,204) sample vectors, each described by 1250 (810) features derived from coincident auxiliary channel information.  
The samples are most representative of the background in gravitational-wave burst searches which generally target short, unmodeled signals.

``Clean'' (Class 0) samples, $\{\auxvec\}_0$, are formed by first generating 10$^5$ randomly distributed times within the data from each detector, producing a Poisson distribution mimicking (at a greatly exaggerated rate) that which would come from a set of true gravitational-wave signals (ignoring effects like detector sensitivity). It can also be seen as a sampling of times when there was no glitch in the gravitational-wave channel, and will sample the typical auxiliary transients that do not produce a gravitational-wave glitch. To further aid in distinguishing times when there is no disturbance, we exclude Class 0 samples which fall within $\pm$100 ms of a Class 1 sample. As with Class 1, the full set of Class 0 samples $\{\auxvec\}_0$ is built from auxiliary channel information nearby each randomly selected time.

%%%%%%%%%%%%%%%%%%%%%%%%%%%%%%%%%%%%%%%%%%%%%%%%%%%%%%%%%%%%%%%%%%%%%
\section{General formulation of the detection problem}
\label{detection_problem}
%%%%%%%%%%%%%%%%%%%%%%%%%%%%%%%%%%%%%%%%%%%%%%%%%%%%%%%%%%%%%%%%%%%%%

The data analysis problem which we address here can be formulated as the robust identification of transient artifacts (glitches) in the gravitational-wave channel based on the information contained in auxiliary detector channels. Clearly, the solution to this problem is directly related to the solution to the ultimate problem of robust detection and classification of gravitational-wave transients in \ac{LIGO} data. The identification of glitches will reduce the non-Gaussian background and improve the sensitivity of gravitational-wave searches.% for which one should rely on information from gravitational-wave channel, knowledge about the waveform of expected signals and cross-correlation of data from different detectors. 
We leave the question of how the results of our current analysis of the auxiliary channels can be incorporated into the search for transient gravitational waves to future work. 

For a given transient event in the gravitational-wave channel, we construct a
feature vector of auxiliary information, $\auxvec$, following the procedure
outlined in Section~\ref{Dataprep}. Our detection problem reduces to binary
prediction on whether this transient is a glitch (Class 1) or a clean sample
(Class 0) based on $\auxvec$ and only $\auxvec$. In feature-space, $\auxvec \in
\Vdata$, this binary decision can be mapped into identifying domains for Class
1 events, $V_1$, and Class 0 events, $V_0$. We call the surface separating the
two regions the decision surface. Unless the two classes are perfectly
separable, which is typically not the case, there is a non-zero probability for
an event of one class to occur in a domain identified with a different class.
In this case, one would like to find an optimal decision surface separating two
classes in such a way that we maximize the probability of finding events of
Class 1 in $V_1$ at fixed probability of miscatagorizing events from Class 0 in
$V_1$.  This essentially minimizes the probability of incorrectly classifying
events.  The former probability, $P_{1}$, represents the probability of glitch
detection which we also call {\em detection efficiency}, and the latter
probability, $P_{0}$, is called the {\em false alarm probability}. This
optimization principle is often referred to as the Neyman-Pearson
criteria~\cite{neyman-1933}. 

The probability of detection and the probability of false alarm can be expressed in terms of conditional probability density functions for the feature vector, $\auxvec$:                  
\begin{subequations}
\label{detfapprobs}
\begin{align}
P_{1} = \int_{\Vdata}\! \Theta\left(f(\auxvec) - \Ft \right)p(\auxvec \given 1) p(1) \diff \auxvec\,,&\\
\label{FAP} P_{0} = \int_{\Vdata} \! \Theta\left(f(\auxvec) - \Ft \right)p(\auxvec \given 0) p(0) \diff \auxvec \,.
 \end{align}
\end{subequations}
Here $p(\auxvec \given 1)$ and $p(\auxvec \given 0)$ define probability density functions for the feature vector in the presence and absence of a glitch in gravitational-wave data, respectively. $p(1)$ and p$(0)$ are prior probabilities for having a glitch or clean data, related to one another via $p(1) + p(0) = 1$. $\Theta\left(f(\auxvec) - \Ft \right)$ defines the region $V_1$ which signifies a glitch in the gravitational-wave data, and $f(\auxvec) =\Ft$ defines the decision surface. $\Ft$ is a threshold parameter, which corresponds to a specific value of the probability of false alarm through (\ref{FAP}).      

The optimal decision surface is found by maximizing the functional
\begin{equation}
\label{Sfunc}
S[f(\auxvec)] = P_{1}[f(\auxvec)] - l_0\left( P_{0}[f(\auxvec)] - \Pt\right)\,,
\end{equation}
where $\Pt$ is a tolerable value for the probability of false alarm and $l_0$ is a Lagrange multiplier. Setting the variation of this functional with respect to $f(\auxvec)$ to zero leads to the condition for the points on the decision surface
\begin{equation}
\label{deccondition}
\frac{p(\auxvec \given 1) p(1)}{p(\auxvec \given 0) p(0)} = \textsc{constant}\,.
\end{equation}
The ratio of conditional probability density functions,
\begin{equation}
\label{lrdef}
\Lambda(x) \equiv \frac{p(\auxvec \given 1)}{p(\auxvec \given 0)}\,,
\end{equation}
is called the likelihood ratio (sometimes also referred to as the Bayes factor). The \textsc{constant} in the optimality condition (\ref{deccondition}) does not carry any special meaning, and the condition can be satisfied if the decision surface is defined as the surface of constant likelihood ratio~\cite{Biswas2012a},
\begin{equation}
\label{decsurface}
f(\auxvec) = \Lambda(\auxvec) = \Ft \,,
\end{equation}
with $\Ft$ set by the probability of false alarm, $\Pt$, through (\ref{FAP}). Actually, the decision surface can be defined by any monotonic function of the likelihood ratio with trivial redefinition of $\Ft$. There is a unique decision surface for each value of $\Pt\in [0,1]$, and we can label decision surfaces by their corresponding values of $\Pt$. 

The optimization of (\ref{Sfunc})  maximizes the detection probability, $P_{1}\rightarrow P_1^{\textsc{opt}}$, for every value of the probability of false alarm, $P_{0} = \Pt$. The curve $P_{1}^{\textsc{opt}}(P_{0})$ is called the Receiver Operating Characteristic (\ac{ROC}) curve. It is a standard measure of any detection algorithm's performance. We can think of optimizing (\ref{Sfunc}) as maximizing the area under the \ac{ROC} curve. For further details on use of the likelihood ratio in the gravitational-wave searches, see~\cite{Biswas2012a,Biswas2012b}. 

Finding the optimal decision surfaces by direct estimation of the conditional probability density functions, $p(\auxvec \given 1)$ and $p(\auxvec \given 0)$, is an extremely difficult task if the feature vector contains more than a few dimensions. For high-dimensional problems, when no parametric model for these probability distributions is known and with a limited number of experimental samples that can be used to estimate these probability density functions, one has to resort to some other method. \ac{MLA} are well-suited for these detection problems. 

In this paper, we consider three popular \ac{MLA}s: \ac{ANN}, \ac{SVM} and \ac{RF}. They differ significantly in their underlying algorithms and their approaches to classification. This allows us to investigate the applicability of different types of \ac{MLA} to glitch identification in the gravitational-wave data. However, all \ac{MLA} require training  samples of events from both Class 1 and Class 0. The \ac{MLA} classifiers use the training sets to find an optimal classification scheme or decision surface. In the limit of infinitely many samples and unlimited computational resources, different classifiers should recover the same theoretical result, the decision surface defined by the constant likelihood ratio~(\ref{decsurface}). To this end, it is critical that classifiers are trained and optimized using criteria consistent with this result. In Appendix~\ref{appendix:fom}, we explore several standard optimization criteria and derive the decision surfaces they generate in this theoretical limit. We find that all of these criteria lead to a decision surface with constant likelihood ratio. In particular, this is true for the fraction of correctly classified events  and the Gini index criteria that are used by  \ac{ANN}\,/\,\ac{SVM} and  \ac{RF}, respectively.

While all classifiers we investigate here should find the same optimal solution with sufficient data, in practice, the algorithms are limited by the finite number of samples in the training sets and by computational cost. The classifiers have to handle a large number of dimensions efficiently, many of which might be redundant or irrelevant. By no means it is clear that the \ac{MLA} classifiers will perform well under such conditions. It is our goal to demonstrate that they do.

We evaluate their performance by computing \ac{ROC} curves. \footnote{More traditional veto approaches to data quality in gravitational-wave searches use another measure of veto quality. For a given veto configuration consisting of a list of disjoint segments of data, the fractional ``dead-time'' is computed from the sum of the durations of all data segments to be vetoed. While not precisely the same, this quantity is related to the probability of false alarm, $P_0$, which accounts only for the fraction of clean data removed from the search. For a typical rate of glitches of ${\sim}0.1$ Hz, the two measures are almost identical in the most relevant region of $P_0 \le
0.01$. Thus, in that interval the \ac{ROC} curves of this paper can be directly compared to the often used figure of merit, efficiency vs.
fractional dead-time. See for example \cite{Smith:2011an}.} These curves, which map the classifiers' overall efficiencies, are objective and can be directly compared. In addition to comparing the \ac{MLA} classifiers to each other, we benchmark  them using \ac{ROC} curves from the \ac{OVL} algorithm~\cite{OVL}. This method constructs segments of data to be vetoed using a hard time window and a threshold on the significance of transients in the auxiliary channels. The veto segments are constructed separately for different auxiliary channels and are applied in the order of decreasing correlation with the gravitational-wave events. By construction, only pairwise correlations between a single auxiliary channel and the gravitational-wave channel are considered by the \ac{OVL} algorithm. These results have a straightforward interpretation and provide a good sanity check. %We also compare performance of the \ac{MLA}s and the \ac{OVL} to the traditionally designed veto segments used in the searches for transient gravitational-waves in the S4 and S6 \ac{LIGO} science runs. 

In order to make the classifier comparison as fair as possible, we train and evaluate their performances using exactly the same data. Furthermore, we use a round-robin procedure for the training-evaluation cycle, which allows us to use all available glitch and clean samples. Samples are randomized and separated into ten equal subsets. To classify events in the $k^{th}$ subset, we use classifiers trained on all but the $k^{th}$ subset. In this way, we ensure that training and evaluation are done with disjoint sets so that any over-training that might occur does not bias our results. 

An MLA classifier's output is called a rank, $\MLArank \in [0,1]$, and a separate rank is assigned to each glitch and clean sample. Higher ranks generally denote a higher confidence that the event is a glitch. A threshold on this rank maps to the probability of false alarm, $P_{0}$, by computing the fraction of clean samples with greater or equal rank. Similarly, the probability of detection or efficiency, $P_1$, is estimated by computing fraction of glitches with greater or equal rank. Essentially, we parametrically define the \ac{ROC} curve, $P_{1}^{\textsc{opt}}(P_{0})$, with a threshold on the classifier's rank. Synchronous training and evaluation of the classifiers allow us to perform a fair comparison and to investigate various ways of combining the outputs of different classifiers. We discuss our findings in detail in Sections~\ref{comparison} and \ref{combining}.
     
%%%%%%%%%%%%%%%%%%%%%%%%%%%%%%%%%%%%%%%%%%%%%%%%%%%%%%%%%%%%%%%%%
\section{Overview of the machine learning algorithms}
\label{pipeline}
%%%%%%%%%%%%%%%%%%%%%%%%%%%%%%%%%%%%%%%%%%%%%%%%%%%%%%%%%%%%%%%%%

In this section, we give a short overview of the basic properties of the classifiers and the tuning procedures used to determine the best performing configurations for each classifier. Throughout this section, we will use the notation $\auxvec_i$ where $i=1, 2, ...N$ to denote the set of $N$ sample feature vectors. Similarly, $y_i$ will denote the actual class associated with the the $i^\mathrm{th}$ sample feature vector, either Class 0 or Class 1. Predictions about a feature vector's class will be denoted by $y(\auxvec_i)$.

%%%%%%%%%%%%%%%%%%%%%%%%%%%%%%%%%%%%%%%
\subsection{Artificial Neural Network}
%%%%%%%%%%%%%%%%%%%%%%%%%%%%%%%%%%%%%%%
An Artificial Neural Network is a machine learning technique based on simulating the data processing in human brains and mimicking biological neural networks \cite{Hastie:2009,HechtNielsen:1989}. %\textit{Originally this has been designed for artificial intelligence since computer always requires precise inputs and follows sequential instructions only while the human brains can perform very complex and fuzzy tasks in a distributed and parallel manner. }
As is well-known, the human brain is composed of a tremendous number of interconnected neurons, with each cell performing a simple task (responding to an input stimulus). However, when a large number of neurons form a complicated network structure, they can perform complex tasks such as speech recognition and decision-making.

A single neuron is composed of dendrites, a cell body, and an axon. When dendrites receive an external stimulus from other neurons, the cell body computes the signal. When the total strength of the stimulus is greater than the synapse threshold, the neuron is fired and sends an electrochemical signal to other neurons through the axon. This process can be implemented with a simple mathematical model including nodes, a network topology and learning rules adopted to a specific data processing task. Nodes are characterized by their number of inputs and outputs (essentially how many other nodes they talk to), and by the connecting weights associated with each input and output. The network topology is defined by the connections between the neurons (nodes). The learning rules prescribe how the connecting weights are initialized and evolve.

There are a large number of \ac{ANN} models with different topologies. For our purpose, we choose to implement the \ac{MLP} model, which is one of the most widely used models. The \ac{MLP} model has input and output layers as well as a few hidden layers in between. The input vector for the input layer is the auxiliary feature vector, $\auxvec$, while the input for hidden layers and the output layer is a combination of the output from nodes in the previous layer. We will call these intermediate vectors $z$ to distinguish them from the full feature vector. Each layer has several neurons which are connected to the neurons in the adjacent layers with individual connecting weights. The initial structure - the number of layers, neurons, and the initial value of connecting weights - is chosen by hand and\,/\,or through an optimization scheme such as a \ac{GA}. %The feature vector, $\auxvec$, is given as input to the neurons in the input layer and the input to the neurons in subsequent layers is a vector constructed from the outputs of the neurons in preceding layers. Layers between the input and output layers are refered to as hidden layers.

When a neuron's input channels receive an external signal exceeding the threshold value set by an activation function, the neuron is fired. This process can be expressed mathematically as:

\begin{eqnarray}
y(z)=f\left(w \cdot z + b\right),
\end{eqnarray}

\noindent where $y(z)$ is the output, $z$ is an input vector, $w$ are connecting weights, $f$ is an activation function, and $b$ is a bias. One may choose the activation function to be either the identity function, the ramp function, the step function, or a sigmoid function.  We use the sigmoid function defined by 

\begin{equation}\label{sigmoid}
%f\left(w \cdot \auxvec + b\right)=\frac{1}{1+\mathrm{e}^{-2s\left(w \cdot \auxvec + b\right)}}\,,
f\left(w \cdot z + b\right)={\left(1+\mathrm{e}^{-2s(w \cdot z + b)}\right)}^{-1}\,.
\end{equation}

We set the activation steepness $s = 0.5$ in hidden layers and $s = 0.9$ at the output layer. There is a single neuron at the output layer, and the value of that neuron's activation function is used as ANN's rank, $\ANNrank$.

The topological parameters determine the number of connection weights, which must be sufficiently large so that \ac{ANN} has enough degrees of freedom to classify a given datum. The network's flexibility depends on the number of connection weights and should be matched with the size of the training sets and the input data's dimensionality. In our work, the numbers of layers and neurons are chosen so that the total number of connection weights is on the order of $10^4$ when using the entire data set. In order to avoid overtraining, we decrease the number of layers and neurons in the runs in which either the dimensionality or the number of training samples are reduced. 
 
The learning scheme finds the optimal connection weights, $w$, in each layer. In this paper we use the improved version of the Resilient back PROPagation (iRPROP) algorithm \cite{IgelHusk:2000}, which minimizes the error between the output value, $y(\auxvec_i)$, and the known value, $y_i$. In this algorithm, the increase (decrease) factor and the minimum (maximum) step-size determine the change in the connection weights, $\Delta w$, at each iteration during the training. In our work, the default values in the \ac{FANN} library \cite{fann} are used for all parameters except the increase factor, which is set to $\eta^{+}=1.001$. The same learning rules were used in all runs. We should note that \ac{ANN} can be optimized in an alternative way, via a \ac{GA} or other similar methods. When using a \ac{GA}, a combined optimization algorithm for topology, feature and weight selection can be applied to improve the performance of \ac{ANN} \cite{KozaRice:1991,DBLP:1992,ChangLipp:1991,Collins90anartificial,Gruau:1992}. We explore these options in a separate publication~\cite{Kim:2012}.

In addition to the choosing the \ac{ANN} configuration parameters, we found that \ac{ANN} requires data pre-processing. In \ac{ANN}, higher absolute values of the input variables have more effect on the output values, so all components of the feature vector, $\auxvec$, are normalized to the range $[0,1]$. To better resolve small $\Delta t$ values, $\Delta t$ is transformed via a logarithmic function before normalization.

\begin{eqnarray}
	\Delta t'=-\mathrm{sign}(\Delta t)\log|\Delta t|.
\end{eqnarray}
This transformation improves \ac{ANN}'s ability to identify glitches, which tend to have smaller values of $\Delta t$. You can find a more detailed description of the procedure for tuning the \ac{ANN} configuration parameters in \cite{Kim:2012}.

%%%%%%%%%%%%%%%%%%%%%%%%%%%%%%%%%%%%%%%%%%%%%%
\subsection{Support Vector Machines}
%%%%%%%%%%%%%%%%%%%%%%%%%%%%%%%%%%%%%%%%%%%%%
A Support Vector Machine is a machine learning algorithm for binary classification on a vector space~\cite{CortesV:1995,cristianini2000introduction}.
It finds the optimal hyperplane that separates the two classes of training samples. This hyperplane is then used as the decision surface in feature space, and classifies events depending on which side of the hyperplane they fall.

As before, let $ \{(\auxvec_i, y_i) \,|\, i=1, 2, ...N \} $ be the training data set, where $\auxvec_i$ is the feature vector of auxiliary transient information near time $t_i$, and $y_i \in \{1, -1 \} $  is a label that marks the sample as either Class 1 or Class 0, respectively. Then assume that the training set is separable by a hyperplane $w \cdot \auxvec-b=0$, where $w$ is the normal vector to the hyperplane and $b$ is the bias. Then the training samples with $y_i = 1$ satisfy the condition $w \cdot \auxvec_i-b \ge 1$, and the training samples with $y_i = -1$ satisfy the condition $w \cdot \auxvec_i-b \le -1$. \ac{SVM} uses a quadratic  programming method to find the $w$ and $b$ that maximize the margin between the hyperplanes $w \cdot \auxvec-b = 1$ and $w \cdot \auxvec-b = -1$.

If the training samples are not separable in the original feature space, $\Vdata$, \ac{SVM} uses a mapping, $\phi(\auxvec)$, into a higher dimensional vector space, $V_{\phi}$, in which two classes of events can be separated. The decision hyperplane in $V_{\phi}$ corresponds to a non-linear surface in the original space, $\Vdata$. Thus, mapping the problem into a higher dimensional space allows \ac{SVM} to consider non-linear decision surfaces. The dimensionality of $V_{\phi}$ grows exponentially with degree of the non-linearity of the decision surfaces in $\Vdata$. As a result, \ac{SVM} can not consider arbitrary decision surfaces and usually has to deal with non-separable populations. If the training samples are not separable after mapping, a penalty parameter, $C$, is introduced to weight the training error. Finding the optimal hyperplane is reduced to the following quadratic programming problem:

\begin{subequations}
\label{SVMprogproblem}
\begin{flalign}
&\min_{w, b, \xi} \left(\frac {1}{2} w \cdot w + C \sum_{i=1}^{N} \xi_i \right)\,, \\
&\text{subject to  } y_i \cdot (w \cdot \phi(\auxvec_i)+b) \ge 1-\xi_i \,, \\
&\xi_i \ge 0, \, i=1, 2, ..., N \,.
 \end{flalign}
\end{subequations}

\noindent When the solution is found, \ac{SVM} classifies a sample $\auxvec$ by the decision function:

\begin{equation}
\label{SVMdecfunction}
y(\auxvec_i) = \mathrm{sign}\left(w \cdot \phi(\auxvec_i) + b \right)\,.
\end{equation}

In solving the quadratic programming problem, the function $\phi$ is not explicitly needed. It is sufficient to specify $\phi(\auxvec_i) \cdot \phi(\auxvec_j)$. The function $K(\auxvec_i, \auxvec_j) = \phi(\auxvec_i) \cdot \phi(\auxvec_j)$ is called the kernel function. The form of the kernel function implicitly  defines the family of surfaces in $\Vdata$ over which {SVM} is optimizing. In this study we use the \ac{RBF} as the kernel function. It is defined as
\begin{equation}
K(\auxvec_i, \auxvec_j) = \mathrm{exp}\left\{-\gamma || \auxvec_i - \auxvec_j ||^2\right\}\,,
\end{equation}
where $\gamma$ is a tunable parameter.

The SVM algorithm was implemented by using the open-source package libsvm~\cite{CC01a}. As part of this package, the kernel function parameter, $\gamma$, and the penalty parameter, $C$, are tuned in order to achieve the best performance for a specific application. The best parameters ($C$ and $\gamma$) are selected through an exhaustive search. For each pair of parameters ($\log C$, $\log \gamma$) on a grid, we calculate a figure of merit. The parameters with the best figure of merit are then used for classifying events. The default figure of merit in libsvm is the accuracy (fraction of correctly classified events). However, we replace it with a figure of merit better adapted to glitch detection. Instead of using the accuracy, our code calculates the area under the estimated \ac{ROC} curve ($P_1^{\textsc{est}}(P_0)$) in the interval of the probability of false alarm $P_0 \in [0.001, 0.05]$ on a log-linear scale ($[0.001, 0.05]$ is a range of acceptable probability of false alarm for practical glitch detection).

\begin{equation}\label{SVM_figure_of_merit}
\mathrm{figure\ of \ merit} = \int\limits^{P_0 = 0.05}_{P_0=0.001} \mathrm{d}\,(\mathrm{ln} P_0)\ P_1^{\textsc{est}}(P_0)
\end{equation}

Performing an exhaustive search for the best SVM parameters is computationally expensive. We can speed up this tuning process by exploiting the fact that the tuning time grows non-linearly with the training sample size. By using smaller training sets, we can reduce the total time spent determining the optimal parameters. We randomly selected $p$ subsets of vectors from the training set, with each subset 10 times smaller than the full training set. The best pair of the SVM parameters for each of the $p$ subsets was then calculated, with each subset optimization running on a single CPU core. This gives $p$ sets of best parameters, calculated in parallel. The parameters $C$ and $\gamma$ that are selected the most often were then chosen as the final best SVM parameters. This modified parameter optimization algorithm was applied to various training sets (described in  Section~\ref{tuning}). We found that the optimal SVM parameters do not depend on the training set. We, therefore, use the same parameters for all calculations reported in this paper ($C=8$ and $\gamma=0.0078125$).

In its standard configuration, SVM classifies samples by a discrete label, $y \in \{1, -1\}$. However, the libsvm package can provide a probability based version of (\ref{SVMdecfunction}) that yields continuous values, $\SVMrank \in [0, 1]$~\cite{Wu:2004}. We use these continuous values as the output of the SVM classifier.

%%%%%%%%%%%%%%%%%%%%%%%%%%%%%%%%%%%%%%%%%%
\subsection{Random Forest Technology}
%%%%%%%%%%%%%%%%%%%%%%%%%%%%%%%%%%%%%%%%%
Random forest technology~\cite{Breiman:1996,Breiman01}  improves upon the classical decision tree~\cite{cart84,Quinlan86} approach to classification. The classifying decision tree performs a series of binary splits on any\,/\,all of the dimensions of the feature vector, $\auxvec$, that describes an event. The goal is to distribute events into groups consisting of only a single class. In a machine-learning context, the decision tree is formed by ``training'' it on a set of events of known class. A series of splits are made, where each split chooses the dimension and threshold that optimizes a certain criterion, such as the fraction of correctly classified training events or the Gini index. Splitting stops once no split can further improve the optimization criterion or the limit on the minimum number of events allowed on a branch (the furthest reaches of a decision tree) is reached; at this point the branch becomes a leaf. Once a tree is formed, an event of unknown class is fed into the tree, and depending on its feature vector, $\auxvec$, it will be labeled as either Class 0 or Class 1. However, a single decision tree can be a victim to both false minima and over-training. To guard against this, we create a forest of decision trees and average over their answers; this results in a continuous ranking, $\RFrank \in [0,1]$, rather than a binary classification, as events can be placed on a continuum between Class 0 and Class 1.

Each decision tree in the forest is trained on a bootstrap replica of the
original training set. If the original training set has N events, each
bootstrap replica will also have N events, which are chosen randomly with
replacement -- so that any given event may be picked more than once. Therefore,
each tree gets a different set of training events. To further avoid false
minima, random forest technology chooses a different random subset of the
features to be available for splitting at each node. This ensures that a
peculiarity in a particular dimension does not dominate the decision making
process.

We use the StatPatternRecognition software package's~\cite{StatPatternRecognitionPackage} implementation of \ac{RF}. The key input parameters are the number of trees in the forest, the number of features randomly selected for splitting at each tree node (branching point), the minimum number of samples on the terminal tree nodes (leaves) and the optimization criterion. To determine the best set of the \ac{RF} parameters, we perform a search over a coarse grid in the parameter space, maximizing efficiency or the detection probability, $P_1$, at the probability of false alarm, $P_0 = 0.01$. We find that beyond a certain point, the \ac{RF} efficiency grows very slowly with the number of trees and the number of features selected for splitting  at the cost of a significant increase in running time during the training process. Taking this into account, we arrive at the following configuration which we use in all runs: 100 trees in the forest, 64 features for splitting, a minimum of 8 samples on a node, and the Gini index as the optimization criterion.

%%%%%%%%%%%%%%%%%%%%%%%%%%%%%%%%%%%%%%%%%%%%%%
\subsection{Ordered Veto List Algorithm}
%%%%%%%%%%%%%%%%%%%%%%%%%%%%%%%%%%%%%%%%%%%%%%%
The \ac{OVL} algorithm operates by looking for coincidences between the transients in gravitational-wave and auxiliary channels. Specifically, the transients identified in the auxiliary channel are used to construct a list of time segments. All transients in the gravitational-wave channel occurring within these segments are removed from the list of transient gravitational-wave candidates. In effect, the data in these time segments are vetoed prior to any search for gravitational-waves.  

The algorithm applies a hierarchical approach, assuming that transients in certain auxiliary channels are more correlated with the glitches in gravitational-wave channel, and looks for a hierarchy of different types of correlations between auxiliary and gravitaitonal-wave glitches. Specifically, a series of veto configurations is created, corresponding to different auxiliary channels, the time windows around transients and the threshold on their significance. The ordered list corresponds to a list of these configurations, and veto configurations are applied to the data in order of decreasing correlation. For this study, the maximum time window is set to $\pm$ 100 ms to match the one we use to create auxiliary feature vectors for the MLA classifiers (Section~\ref{Dataprep}). Similarly, the lowest threshold on significance, $\rho$, is set to the auxiliary channel nominal threshold of 15. For each channel, the number of possible veto configurations is equal to the number of unique combinations that can be constructed from the list of the time windows, [$\pm$ 25 ms, $\pm$ 50 ms, $\pm$ 100 ms], and  the significance thresholds, [15, 25, 30, 50, 100, 200, 400, 800, 1600]. 

Importantly, a segment removed by a veto configuration is not seen by later configurations. This prohibits duplicate vetoes and results in a measurement of the additional information contained in subsequent veto configurations. The performance of each configuration is evaluated and they are re-ranked accordingly. The \ac{OVL} algorithm defines the veto-configuration rank, $\OVLrank$, as the ratio of the fraction of gravitational-wave glitches removed to the fraction of analysis time removed. Repeated application of the algorithm produces an ordered list with the best performing configurations appearing first. 

Only some of the veto configurations make it to the final list. Those which perform poorly ($\OVLrank \le 3$) are discarded. This is done in order to get rid of irrelevant or redundant channels and to speed up the algorithm's convergence. Typically, the \ac{OVL} algorithm converges within less than 10 iterations to a final ordered list. We find that only 47 out of 162 auxiliary channels in S4 data and 35 out of 250 auxiliary channels in S6 data appear on the final list. Below, we refer to this subset of channels as the ``{OVL} auxiliary channels.'' For a more detailed description of the OVL algorithm, see \cite{OVL}.

The procedure for optimizing the ordered list of veto configurations can be considered as training phase. An ordered list of veto
configurations, optimized  for a given segment of data, can be  applied
to another segment of data. Veto segments are generated based on the
transients in the auxiliary channels and the list of configurations.
Performance of the algorithm is evaluated by counting fractions of
removed glitches and clean samples, and computing the \ac{ROC} curve. Similarly to \ac{MLA} classifiers
we use the round-robin procedure for \ac{OVL}'s training-evaluation cycle.
%%%%%%%%%%%%%%%%%%%%%%%%%%%%%%%%%%%%%%%%%%%%%%%%%%%%%%%%%%%%%%%%%%%%%%%%%%%%%%%%%%%%
\section{Testing the algorithms' robustness} % against input data variations}
\label{tuning}
%%%%%%%%%%%%%%%%%%%%%%%%%%%%%%%%%%%%%%%%%%%%%%%%%%%%%%%%%%%%%%%%%%%%%%%%%%%%%%%%%%%%%%%

One of the main goals of this study is to establish if \ac{MLA} methods can successfully identify transient instrumental and environmental artifacts in \ac{LIGO} gravitational-wave data. The potential difficulty arises from the high dimensionality and the fact that information from a large number of dimensions might be either redundant or irrelevant. Furthermore, the origin of a large fraction of glitches is unknown in the sense that their cause and effect have not been pinpointed to a single instrumental or environmental source. In the absence of such deterministic knowledge, one has to monitor a large number of auxiliary channels and look for statistically significant correlations between transients in these channels and transients in the gravitational-wave channel. These correlations, in principle, may involve more than one auxiliary channel and may depend on the transients' parameters in an extremely complicated way. Additionally, new kinds of artifacts may arise if one of the detector subsystems begin to malfunction. Likewise, some auxiliary channels' coupling strengths to the gravitational-wave channel may be functions of the detector's state (e.g. optical cavity configuration and mirror alignment). Depending on the detector's state, the same disturbance witnessed by an auxiliary channel may or may not cause a glitch in the gravitational-wave channel. This information can not be captured by the Kleine Welle-derived parameters of the transients in the auxiliary channels alone and requires extending the current method. We leave this problem to future work.  

Because of the uncertainty in the types and locations of correlations, we include as many auxiliary channels and their transients' parameters as possible. However, this forces us to handle a large number of features, many of which might be either redundant or irrelevant. The \ac{MLA} classifiers may be confused by the presence of these superfluous features and their performance may suffer. One can improve performance by reducing the number of features and keeping only those that are statistically significant. However, this requires pre-processing the input data and tuning, which may be extremely labor intensive. On the other hand, if the \ac{MLA} classifier can ignore irrelevant dimensions automatically without a significant decrease in performance, it can be used as a robust analysis tool for real-time glitch identification and detector characterization. By efficiently processing information from all auxiliary channels, a classifier will be able to identify new artifacts and help to diagnose problems with the detector.

In order to determine our classifiers' robustness, we perform a series of runs in which we vary the dimensionality of the input data and evaluate the classifiers' performance. First, we investigate how their efficiency depends on which transient parameters are used. We expect that not all of the five parameters, ($\rho$, $\Delta t$, $f$, $d$, $n$), are equally informative. Naively, $\rho$ and $\Delta t$, reflecting the disturbance's amplitude in the auxiliary channel and its degree of coincidence with the transient in gravitational-wave channel, respectively, should be the most informative. Potentially, the frequency, $f$, duration, $d$, and the number of wavelet coefficients, $n$, may carry useful information if only certain auxiliary transients produce glitches. However, it is more likely that these parameters are correlated with the corresponding parameters of gravitational-wave transient, which we do not incorporate in this analysis. Such correlations, even if not broadened by frequency dependent transfer functions, would require analysis specialized to specific gravitational-wave signals and goes beyond the scope of this paper. We perform a generic analysis, not relying on the specific characteristics of the gravitational-wave transients. 

\begin{figure*}
\centering
\subfloat[S4 \ac{ANN}]{\label{fig:S4_ANN_var_params}\includegraphics[width=0.3\textwidth]{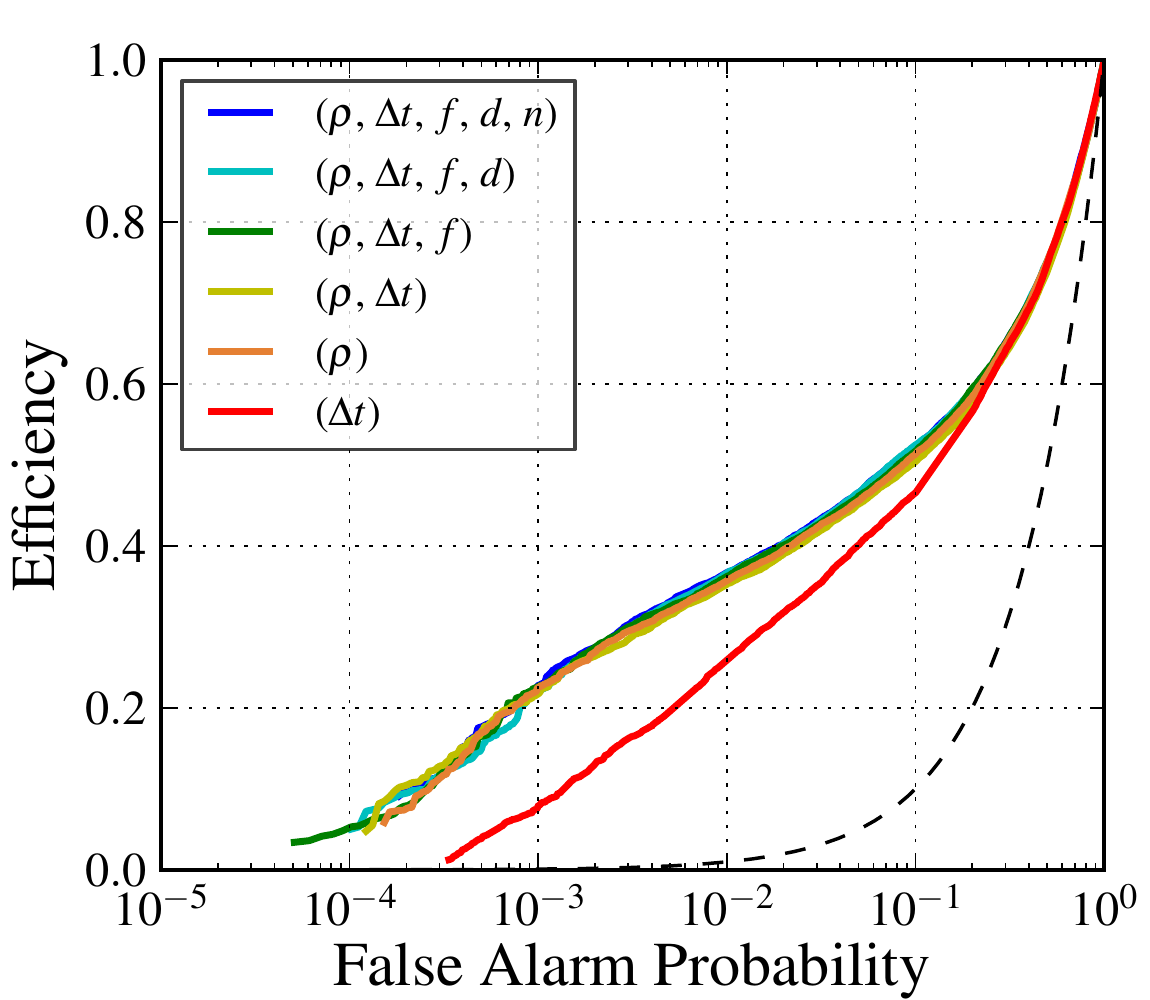}}
\subfloat[S4 \ac{SVM}]{\label{fig:S4_SVM_var_params}\includegraphics[width=0.3\textwidth]{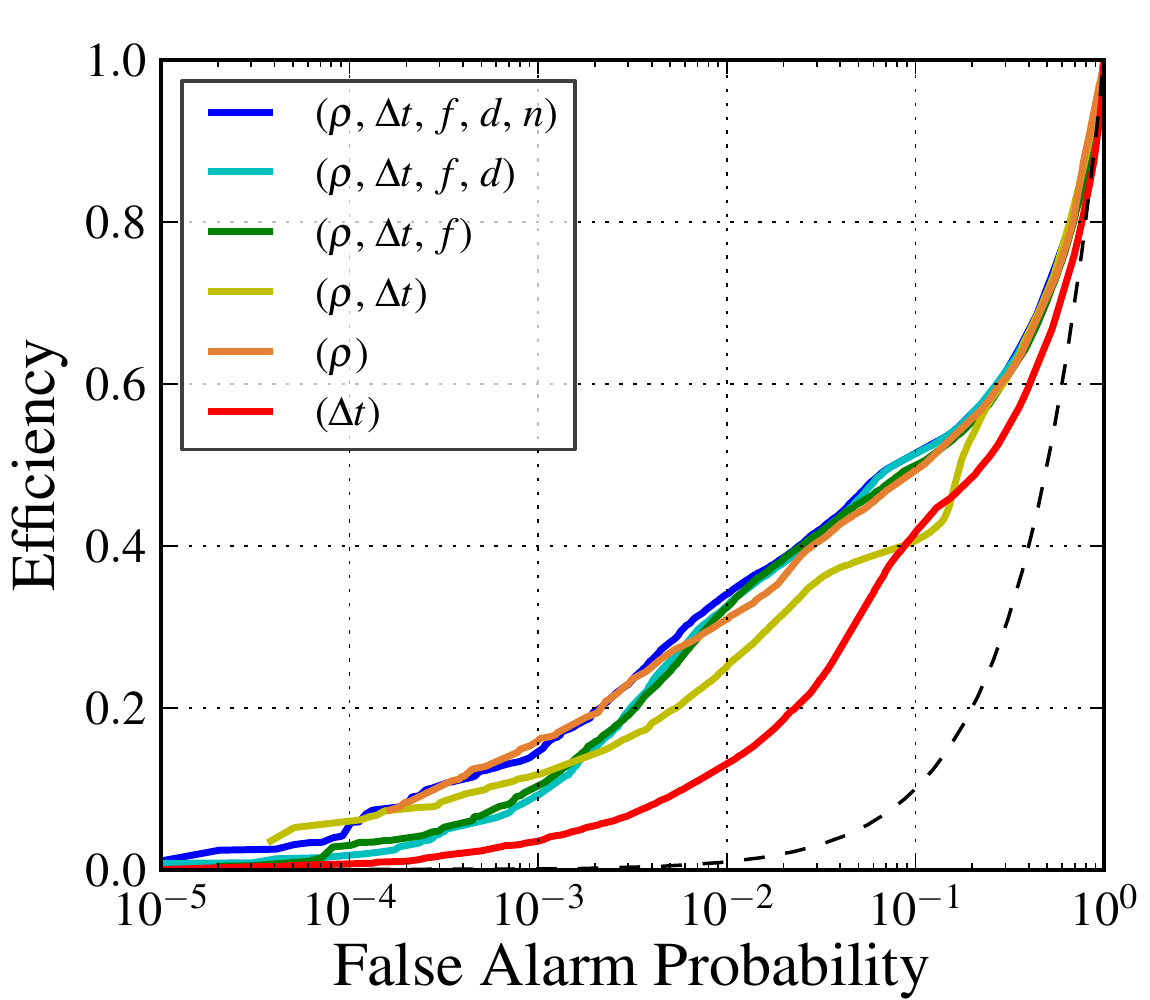}}
\subfloat[S4 \ac{RF}]{\label{fig:S4_MVSC_var_params}\includegraphics[width=0.3\textwidth]{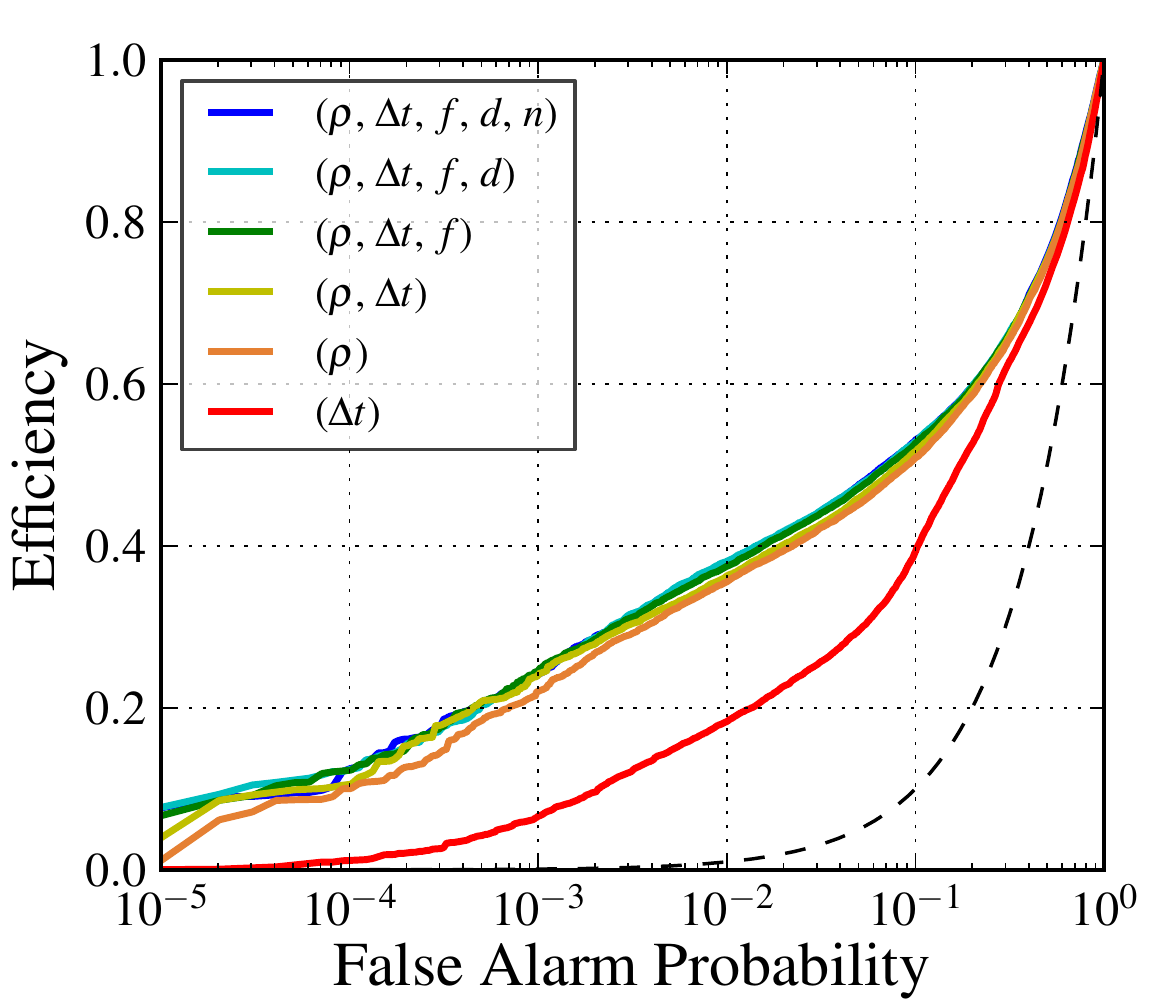}}\\
\subfloat[S6 \ac{ANN}]{\label{fig:S6_ANN_var_params}\includegraphics[width=0.3\textwidth]{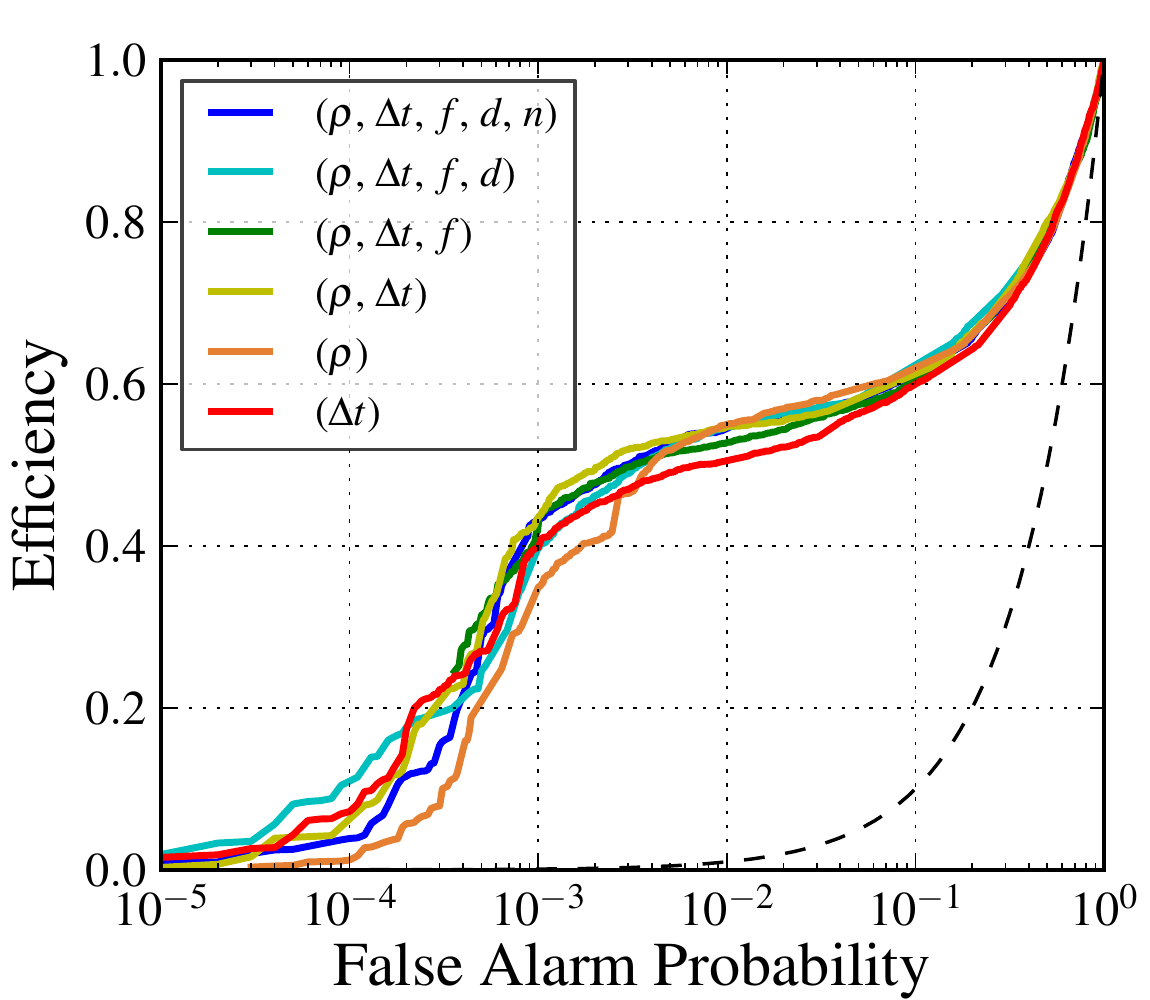}}
\subfloat[S6 \ac{SVM}]{\label{fig:S6_SVM_var_params}\includegraphics[width=0.3\textwidth]{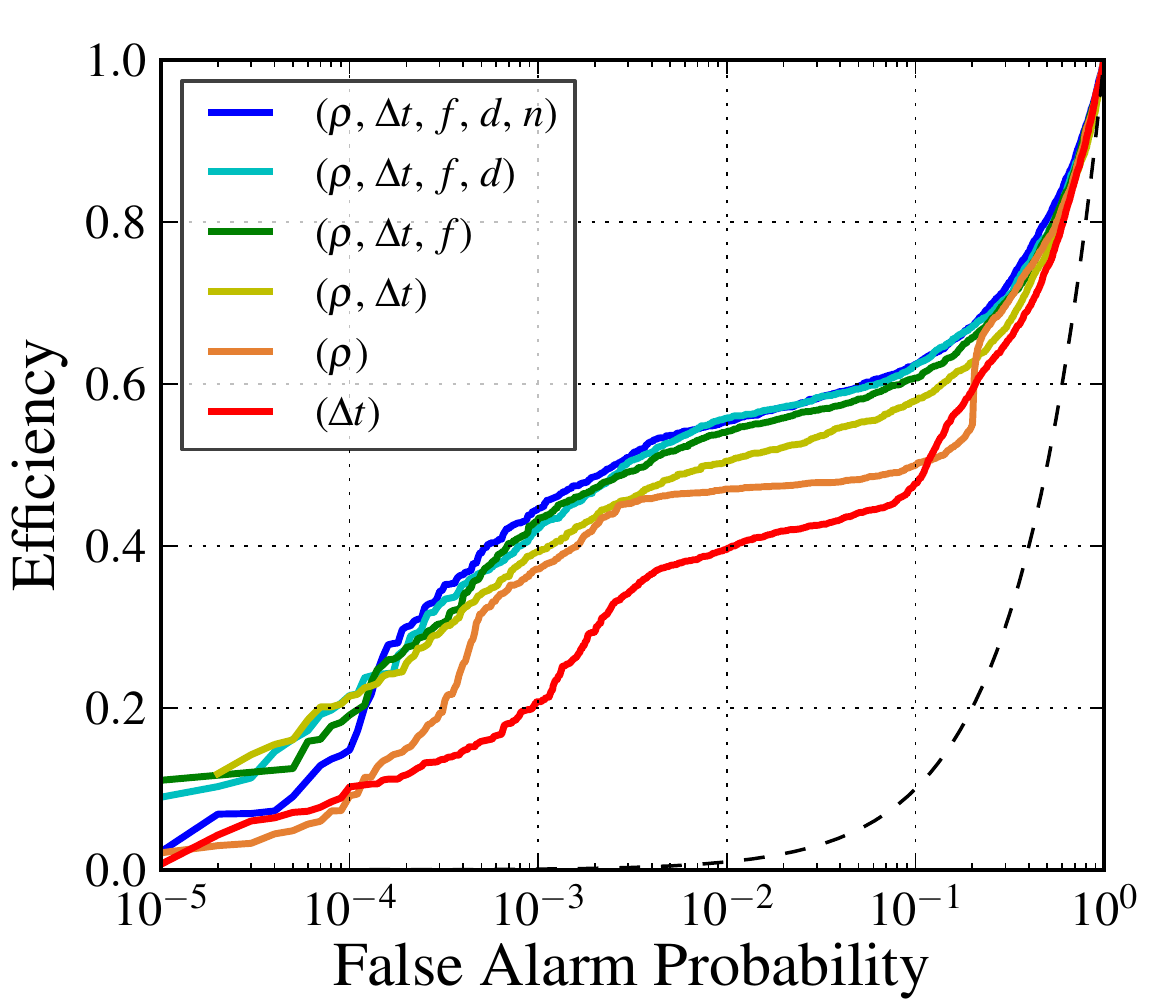}}
\subfloat[S6 \ac{RF}]{\label{fig:S6_MVSC_var_params}\includegraphics[width=0.3\textwidth]{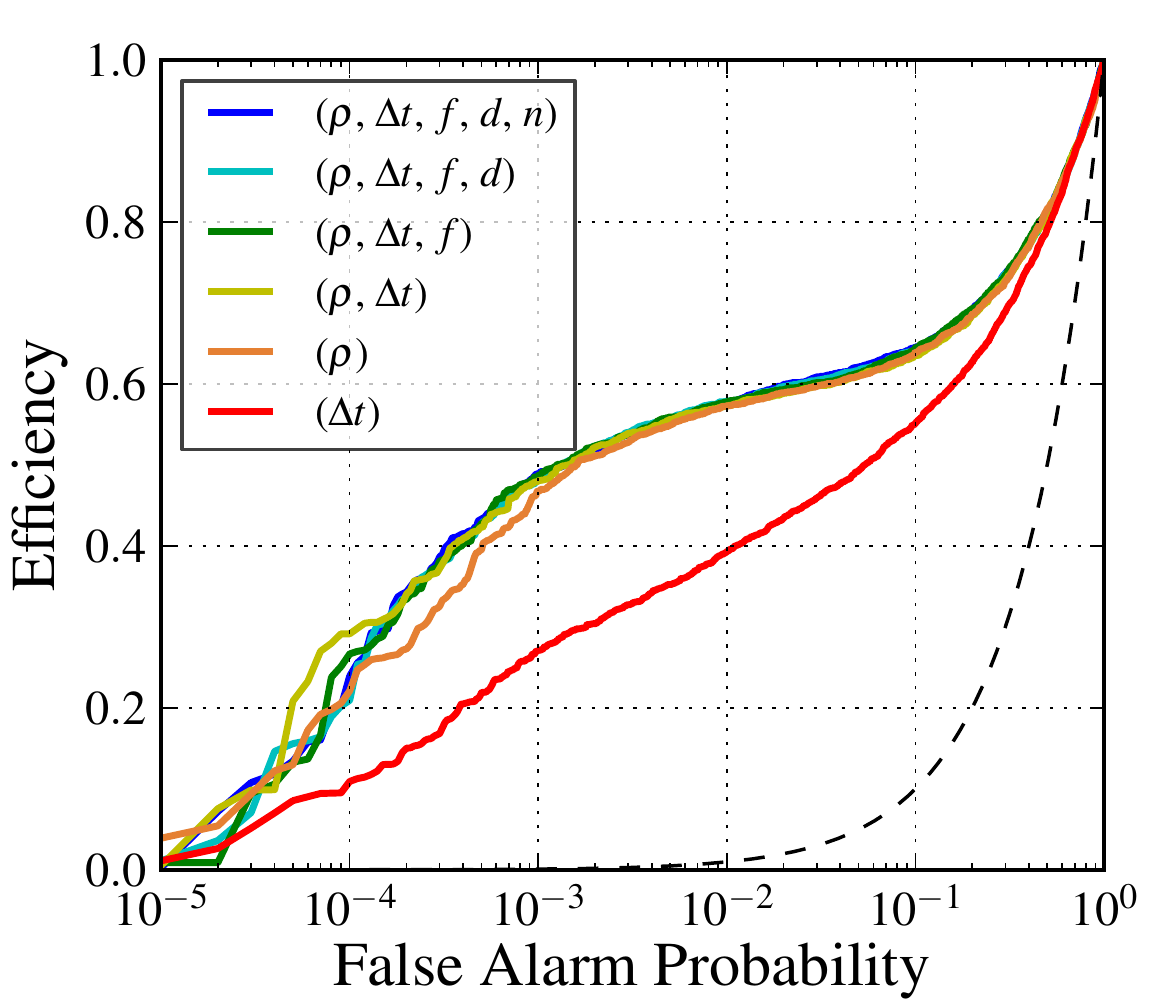}}
\caption{Varying sample features. We expect some of the five features recorded for each auxiliary channel to be more useful than others. To quantitatively demonstrate this, we train and evaluate our classifiers using subsets of our sample data, with each subset restricting the number of auxiliary features. We observe the general trend that the significance, $\rho$, and time difference, $\Delta t$, are the two most important features. Between those two, $\rho$ appears to be marginally more important than $\Delta t$. On the other hand, the central frequency, $f$, the duration, $d$, and the number of wavelet coefficients in an event, $n$, all appear to have very little affect on the classifier's performance. Importantly, our classifiers are not impaired by the presence of these superfluous features and appear to robustly reject irrelevant data without significant efficiency loss. The black dashed line represents a classifiers based on random choice.}
\label{fig:var_params}
\end{figure*}

Anticipating that some of the parameters could be irrelevant, we prepare several data sets by removing features from the list: ($\rho$, $\Delta t$, $f$, $d$, $n$). We prepare these data sets for both S4 and S6 data and run each of the classifiers through the training-evaluation round-robin cycles described in Section ~\ref{detection_problem}. We evaluate their performance by computing the \ac{ROC} curves, shown in \Cref{fig:var_params}.

% including set of figures for variation of trigger parameters
\graphicspath{{figures/}}

We note the following relative trends in the \ac{ROC} curves for all classifiers. The omission of the transient's duration, $d$, and the number of wavelets, $n$, has virtually no effect on efficiency. The \ac{ROC} curves are the same to within our error, which is less than $\pm$ 1 \% for our efficiency measurement (This is based on the total number of glitch samples and the normal approximation for binomial confidence interval, $\sqrt{P_{1}(1-P_{1})/N}$). Omission of the frequency, $f$, slightly reduces the efficiency of \ac{SVM} (\Cref{fig:S4_SVM_var_params} and \Cref{fig:S6_SVM_var_params}), but has no effect on either \ac{ANN} or \ac{RF}. A comparison between the \ac{ROC} curves for ($\rho$, $\Delta t$), ($\rho$) and ($\Delta t$) data sets shows that while a transient's significance is the most informative parameter, including the time difference generally results in better overall performance. Of the three \ac{MLA} classifiers, \ac{SVM} seems to be the most sensitive to whether the time difference is used in addition to significance. \ac{RF}, as it appears, relies primarily on significance, which is reflected in poor performance of the ($\Delta t$)-only \ac{ROC} curves in \Cref{fig:S4_MVSC_var_params} and \Cref{fig:S6_MVSC_var_params}. The trend for \ac{ANN} is not as clear. In S4 data, including timing does not change the \ac{ROC} curve (\Cref{fig:S4_ANN_var_params}) while in S6 data it improves it (\Cref{fig:S6_ANN_var_params}). Overall, we conclude that based on these tests, most if not all the information about detected glitches is contained in $\rho$ and $\Delta t$ pair. At the same time, keeping irrelevant features does not seem to have a negative effect on our classifiers' performance.  

The \ac{OVL} algorithm, which we use as a benchmark, ranks and orders the auxiliary channels based on the strength of correlations between transient disturbances in the auxiliary channels and glitches in gravitational-wave channel. The final list of \ac{OVL} channels includes only a small subset of the available auxiliary channels, 47 (of 162) in S4 data and 35 (of 250) in S6 data. The rest of the channels do not show statistically significant correlations. It is possible that these channels contain no useful information for glitch identification, or that one has to include correlations involving multiple channels and/or other features. In the former case, throwing out irrelevant channels will significantly decrease our problem's dimensionality and may improve the classifiers' efficiency. In the latter case, classifiers might be capable of using higher order correlations to identify new classes of glitches missed by \ac{OVL}. 

\begin{figure*}
\centering
\subfloat[S4 \ac{ANN}]{\label{fig:S4_ANN_reduc_chan}\includegraphics[width=0.3\textwidth]{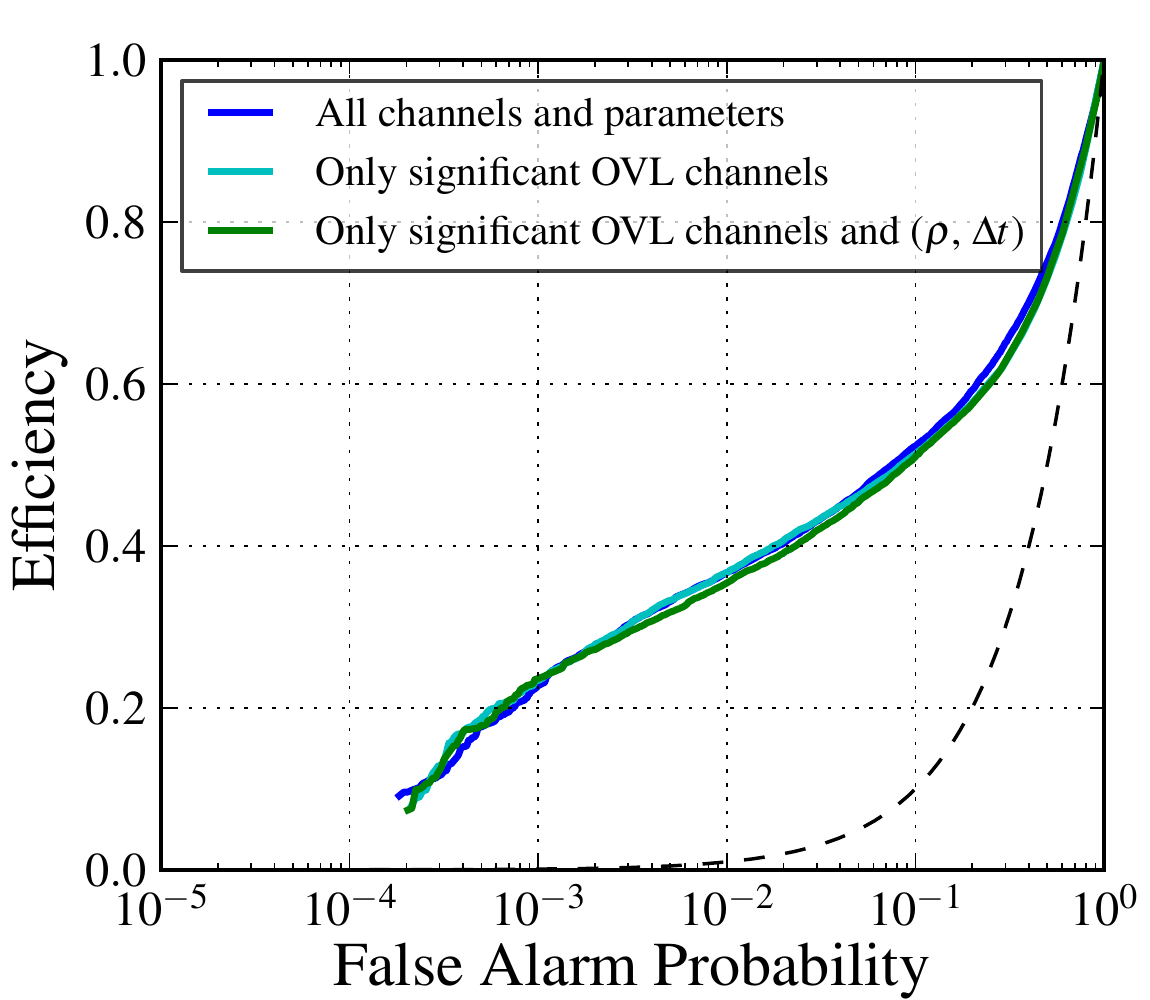}}
\subfloat[S4 \ac{SVM}]{\label{fig:S4_SVM_reduc_chan}\includegraphics[width=0.3\textwidth]{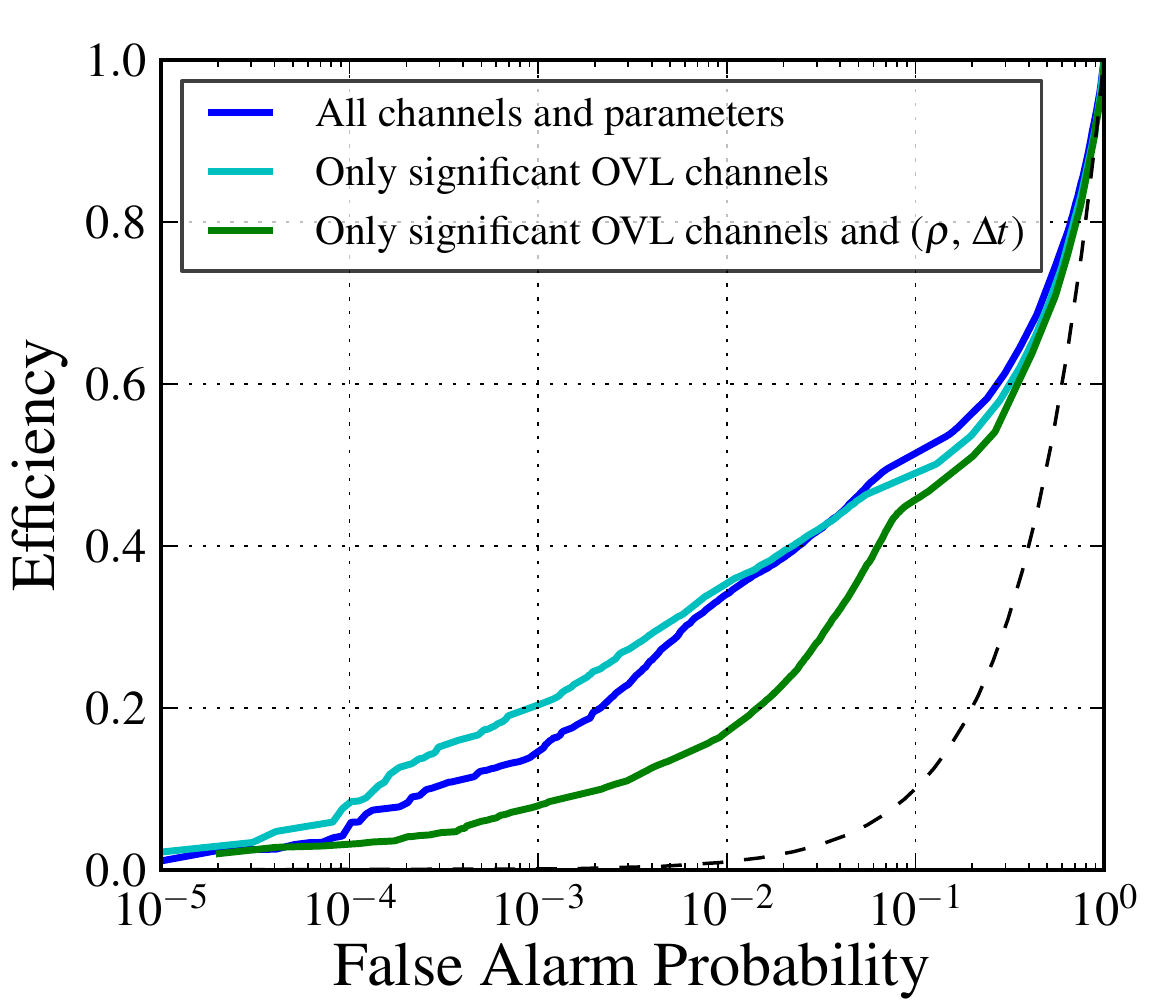}}
\subfloat[S4 \ac{RF}]{\label{fig:S4_MVSC_reduc_chan}\includegraphics[width=0.3\textwidth]{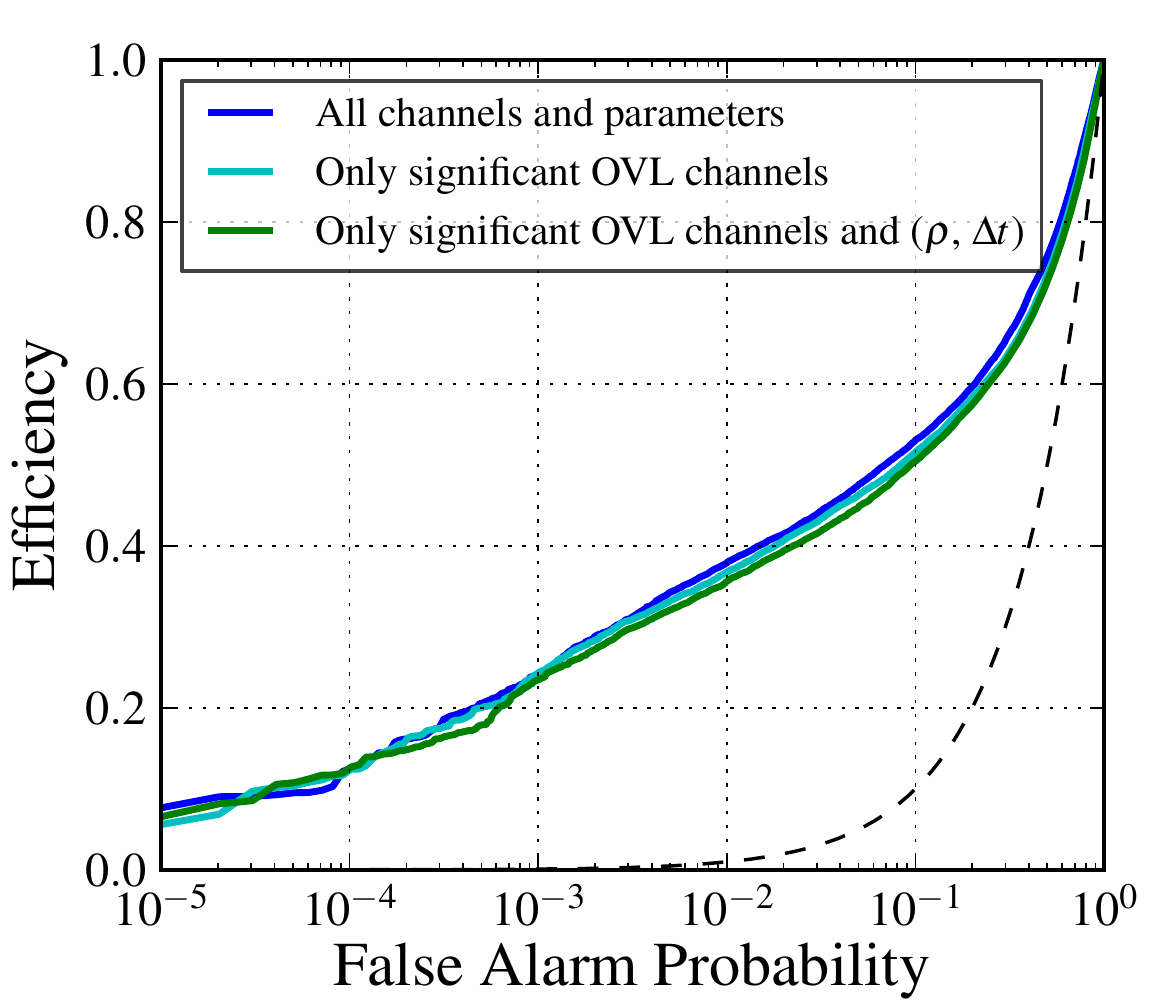}}\\
\subfloat[S6 \ac{ANN}]{\label{fig:S6_ANN_reduc_chan}\includegraphics[width=0.3\textwidth]{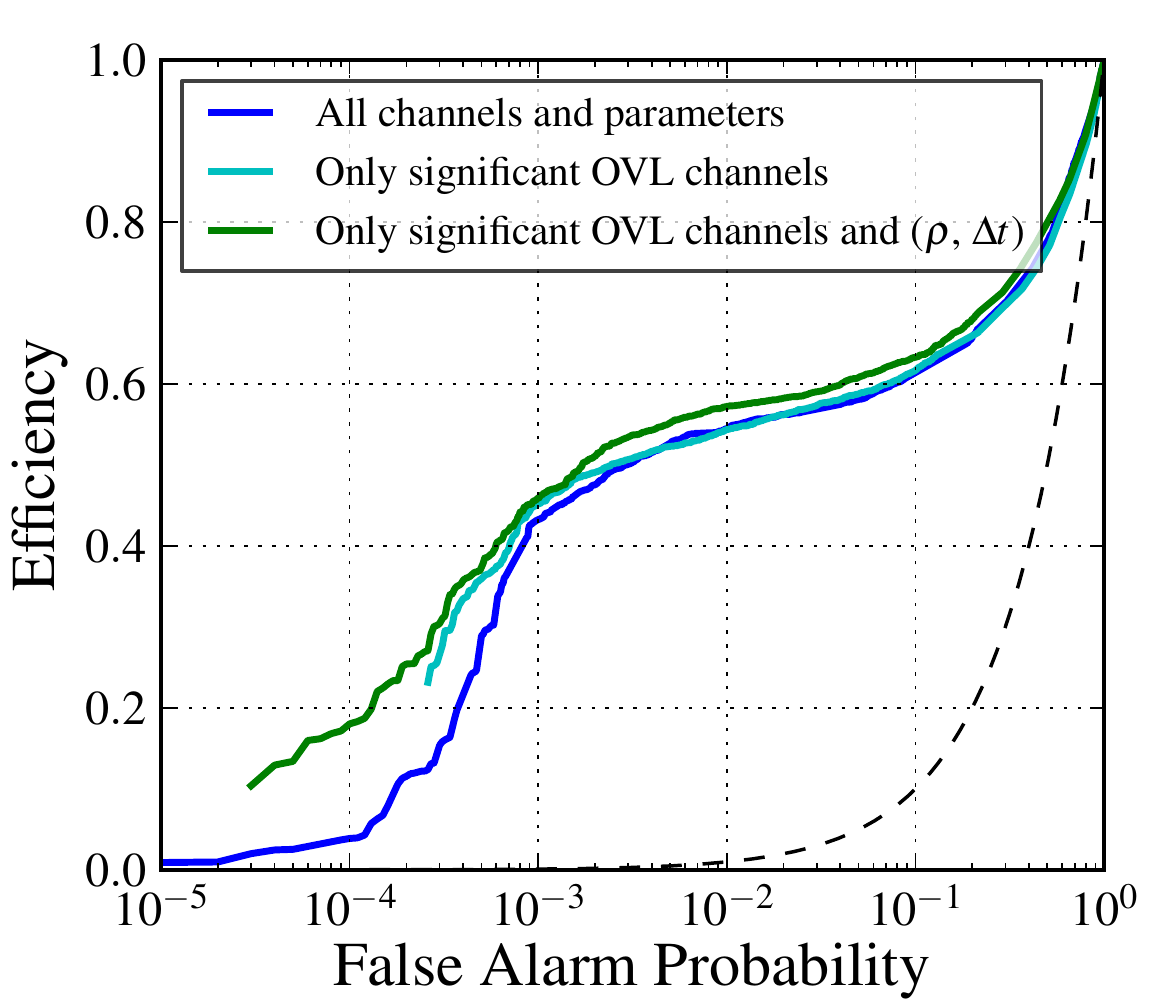}}
\subfloat[S6 \ac{SVM}]{\label{fig:S6_SVM_reduc_chan}\includegraphics[width=0.3\textwidth]{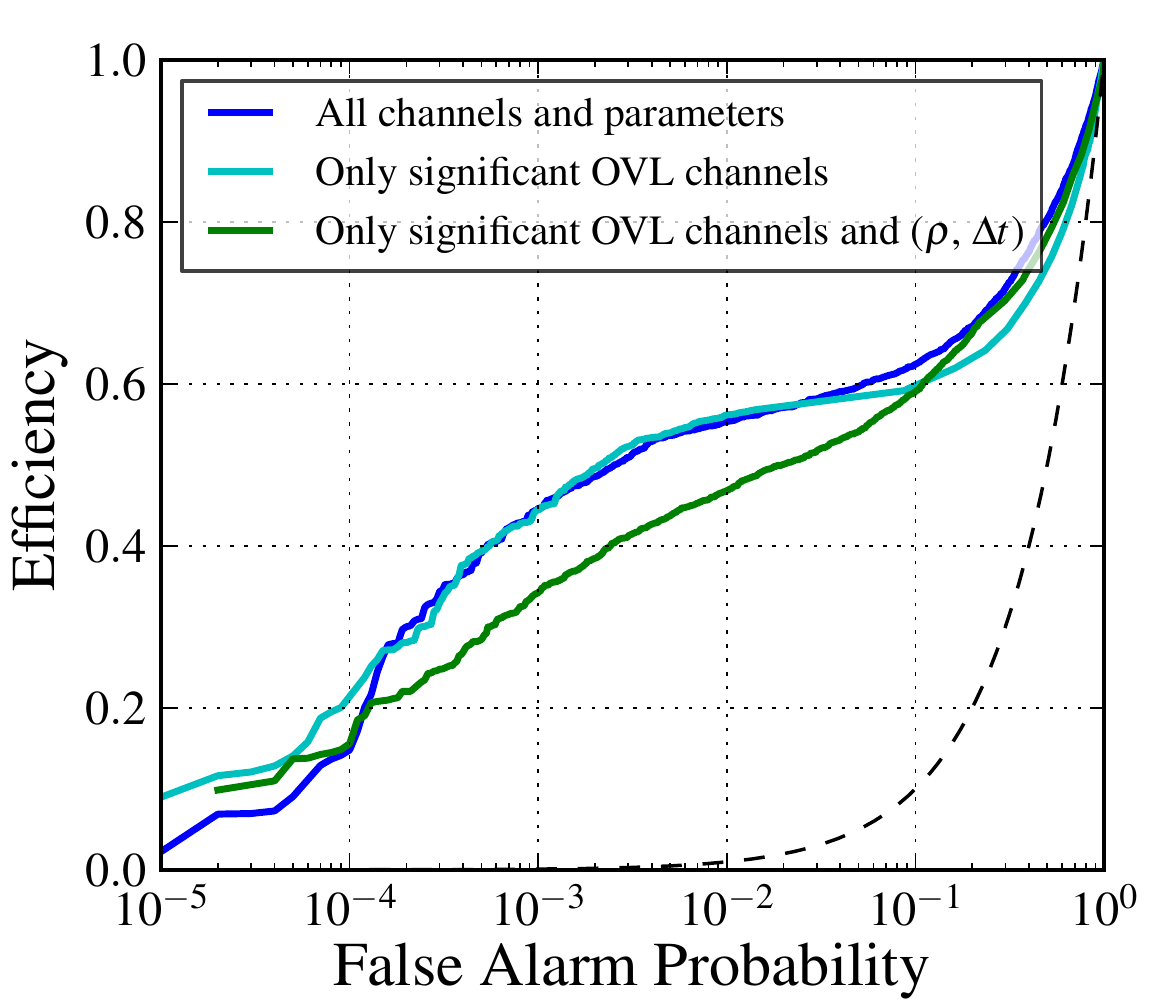}}
\subfloat[S6 \ac{RF}]{\label{fig:S6_MVSC_reduc_chan}\includegraphics[width=0.3\textwidth]{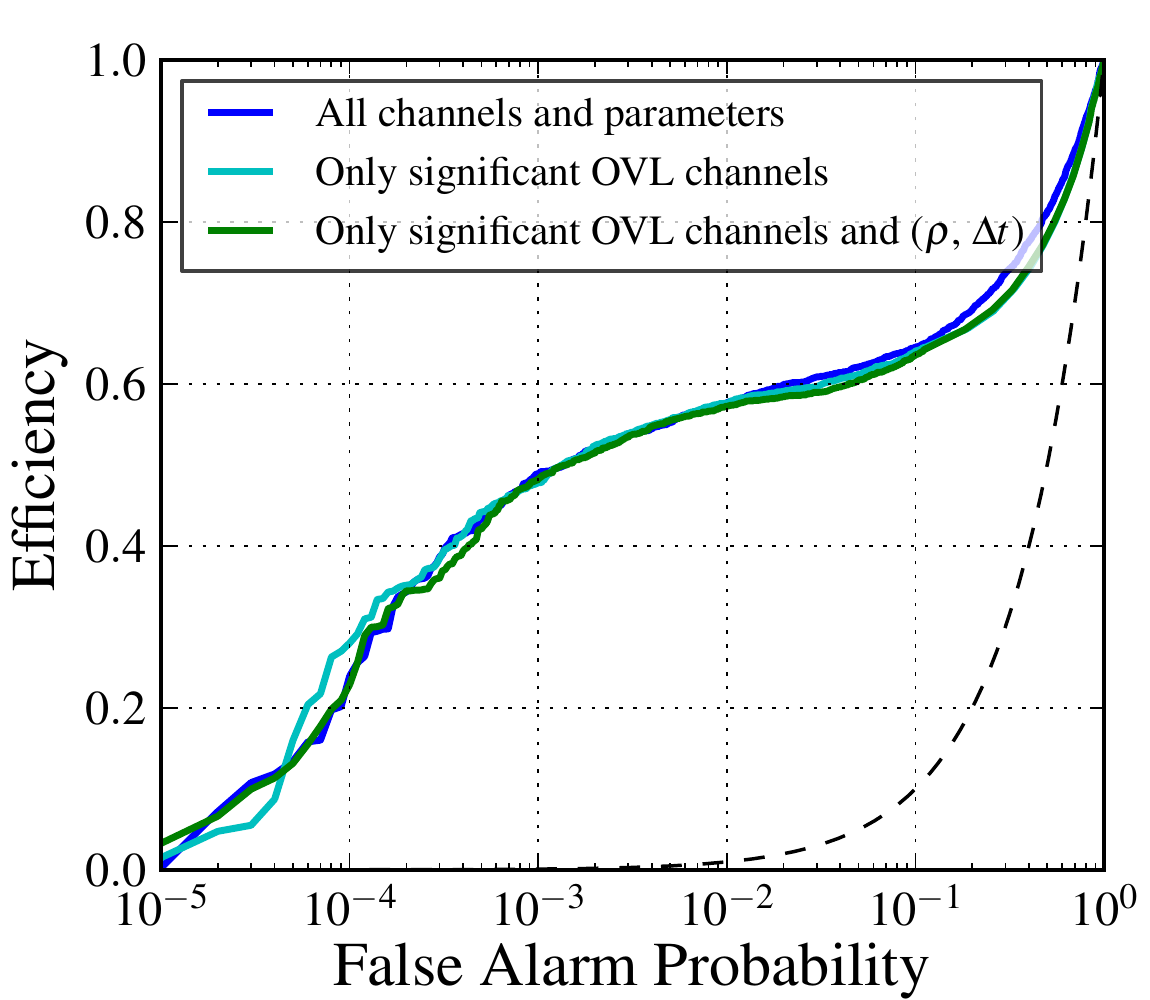}}
\caption{Reducing the number of channels. One way to reduce the dimensionality of our feature space is to reduce the number of auxiliary channels used to create the feature vector. We use a subset of auxiliary channels identified by \ac{OVL} as strongly correlated with glitches in the gravitational-wave channel (light blue). We notice that for the most part, there is not much efficiency loss when restricting the feature space in this way. This also means that very little information is extracted from the other auxiliary channels. The classifiers can reject extraneous channels and features without significant loss or gain of efficiency. We also restrict the feature vector to only include the significance, $\rho$, and the time difference, $\Delta t$, for the OVL auxiliary channels (green). Again, there is not much efficiency loss, suggesting that these are the important features and that the classifiers can robustly reject unimportant features automatically. The black dashed line represents a classifier that is based on random choice.}
\label{fig:reduc_chan}
\end{figure*}

We prepare two sets of data to investigate these possibilities. In the first data set, we use only the {OVL} auxiliary channels and exclude information from all other channels. In the second data set, we further reduce the number of dimensions by using only $\rho$ and $\Delta t$. We apply classifiers to both data sets, evaluate their performance and compare it to the run over the full data set (all channels and all features). \Cref{fig:reduc_chan} shows the \ac{ROC} curves computed for these test runs. 

% including set of figures for reduction of number of channels 

In both S4 and S6 data, the three curves for \ac{RF} (\Cref{fig:S4_MVSC_reduc_chan} and \Cref{fig:S6_MVSC_reduc_chan}) lay on the top of each other, demonstrating that the classifier's performance is not affected by the data reduction. \ac{ANN} shows slight improvement in its performance for the maximally reduced data set in the S6 data (\Cref{fig:S6_ANN_reduc_chan}), and no discernible change in the S4 data (\Cref{fig:S4_ANN_reduc_chan}). \ac{SVM} exhibits the most variation of the three classifiers. While dropping the auxiliary channels not included in the \ac{OVL} list has a very small effect on \ac{SVM}'s ROC curve, further data reduction leads to an efficiency loss (\Cref{fig:S4_SVM_reduc_chan} and \Cref{fig:S6_SVM_reduc_chan}). Viewed together, the plots in \Cref{fig:reduc_chan} imply that, on one hand, non-\ac{OVL} channels can be safely dropped from the analysis, but on the other hand, the presence of these uninformative channels does not reduce our classifiers' efficiency. This is reassuring. As previously mentioned, one would like to use these methods for real-time classification and detector diagnosis, in which case monitoring as many channels as possible allows us to identify new kinds of glitches and potential detector malfunctions. For example, an auxiliary channel that previously showed no sign of a problem may begin to witness glitches. If excluded from the analysis based on its previous irrelevance, the classifiers would not be able to identify glitches witnessed by this channel or warn of a problem.

Another way in which input data may influence a classifier's performance is by limiting the number of samples in the training set. Theoretically, the larger the training sets, the more accurate a classifier's prediction. However, larger training sets come with a much higher computational cost and longer training times. In our case, the size of the glitch training set is limited by the glitch rate in the gravitational-wave channel and the duration of the detector's run. We remind the reader that we use four weeks from the S4 run from the H1 detector and one week from the S6 run from the L1 detector to collect glitch samples. One would like to use shorter segments to better capture non-stationarity of the detector's behavior. However, having too few glitch samples would not provide a classifier  with enough information. Ultimately, the size of the glitch training set will have to be tuned based on the detector's behavior. We have much more control over the size of the clean training set, which is based on completely random times when the detector was operating in the science mode. In our simulations, we start with $10^5$ clean samples, but it might be possible to reduce this number without loss of efficiency, thereby speeding up classifier training. 

\begin{figure*}
\centering
\subfloat[S4 \ac{ANN}]{\label{fig:S4_ANN_var_train_data}\includegraphics[width=0.3\textwidth]{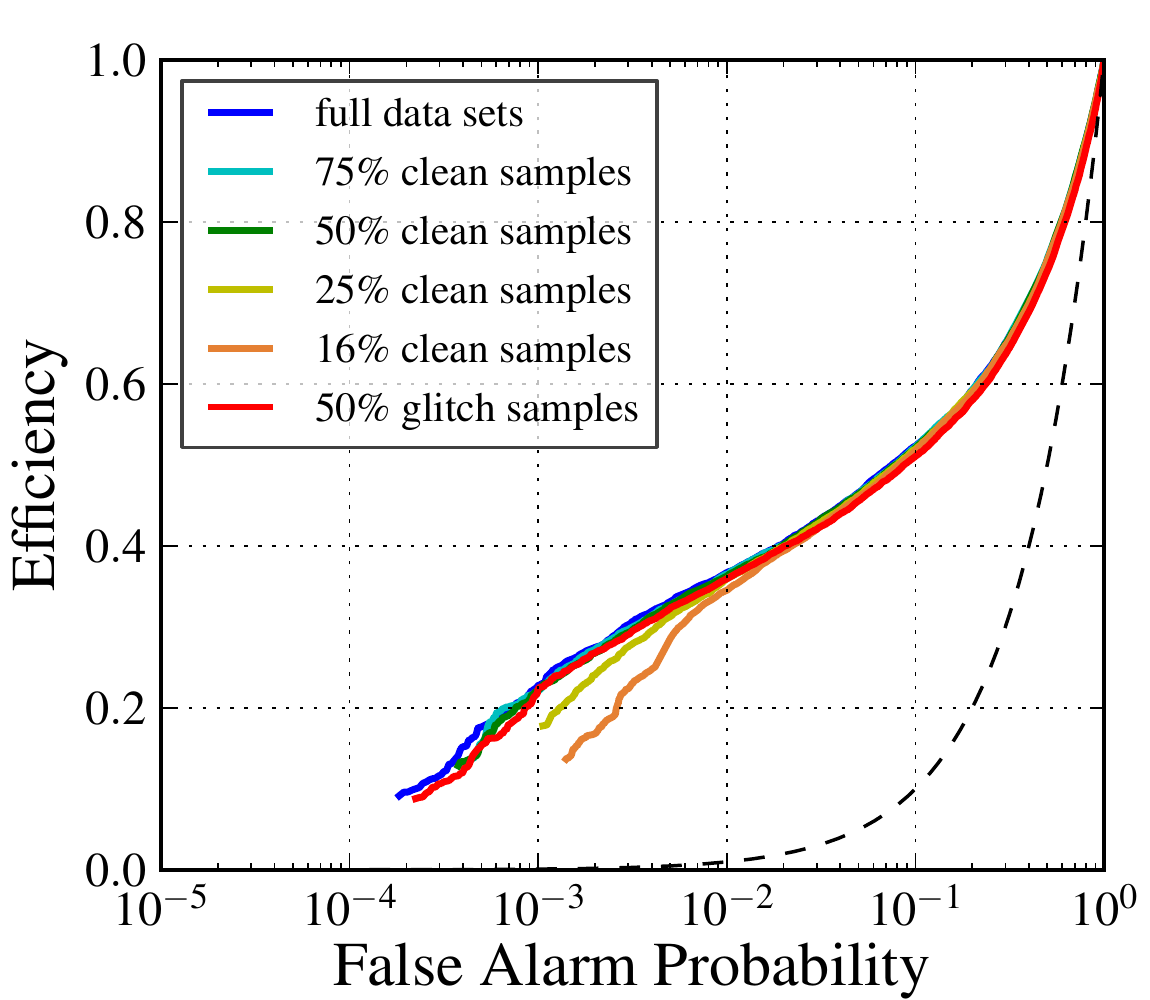}}
\subfloat[S4 \ac{SVM}]{\label{fig:S4_SVM_var_train_data}\includegraphics[width=0.3\textwidth]{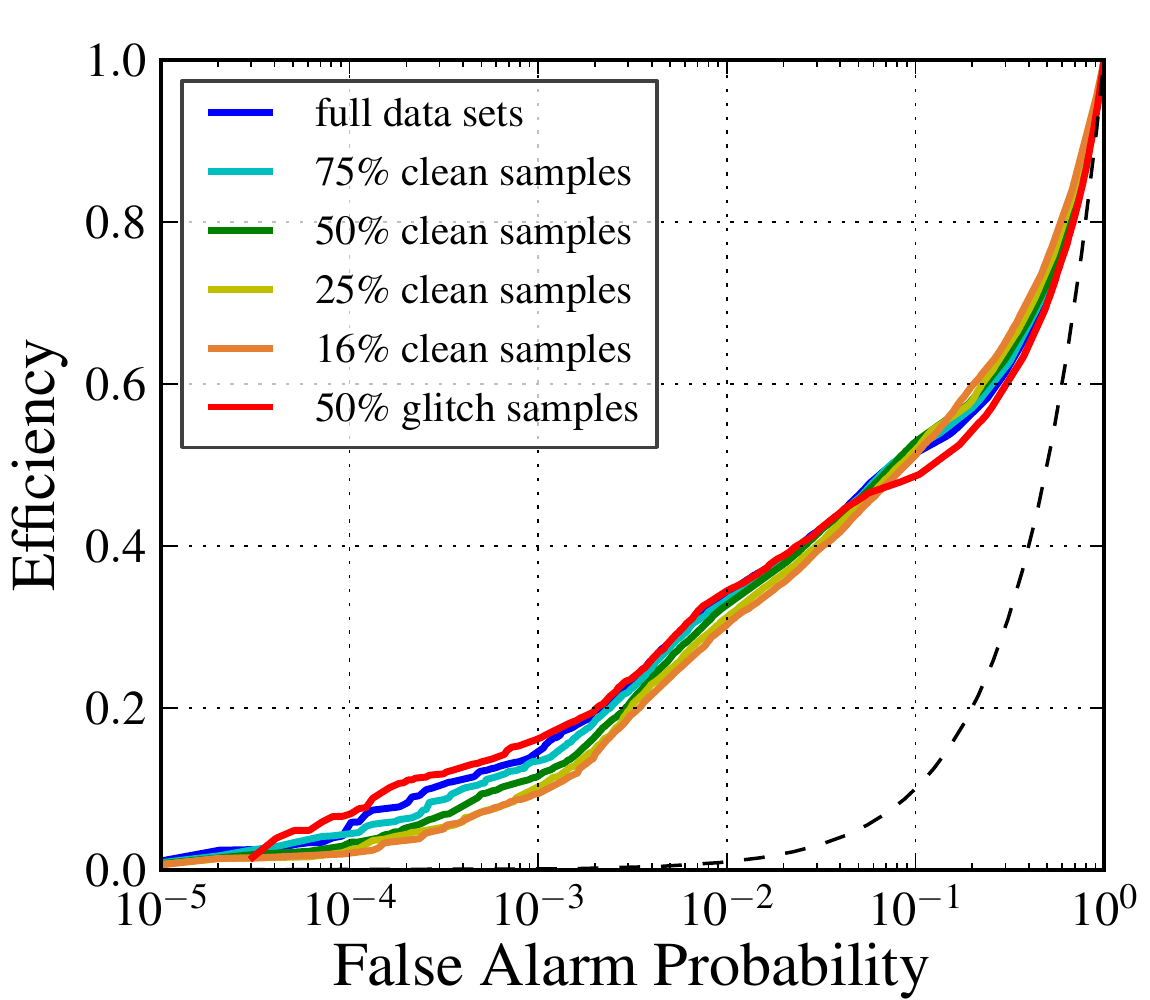}}
\subfloat[S4 \ac{RF}]{\label{fig:S4_MVSC_var_train_data}\includegraphics[width=0.3\textwidth]{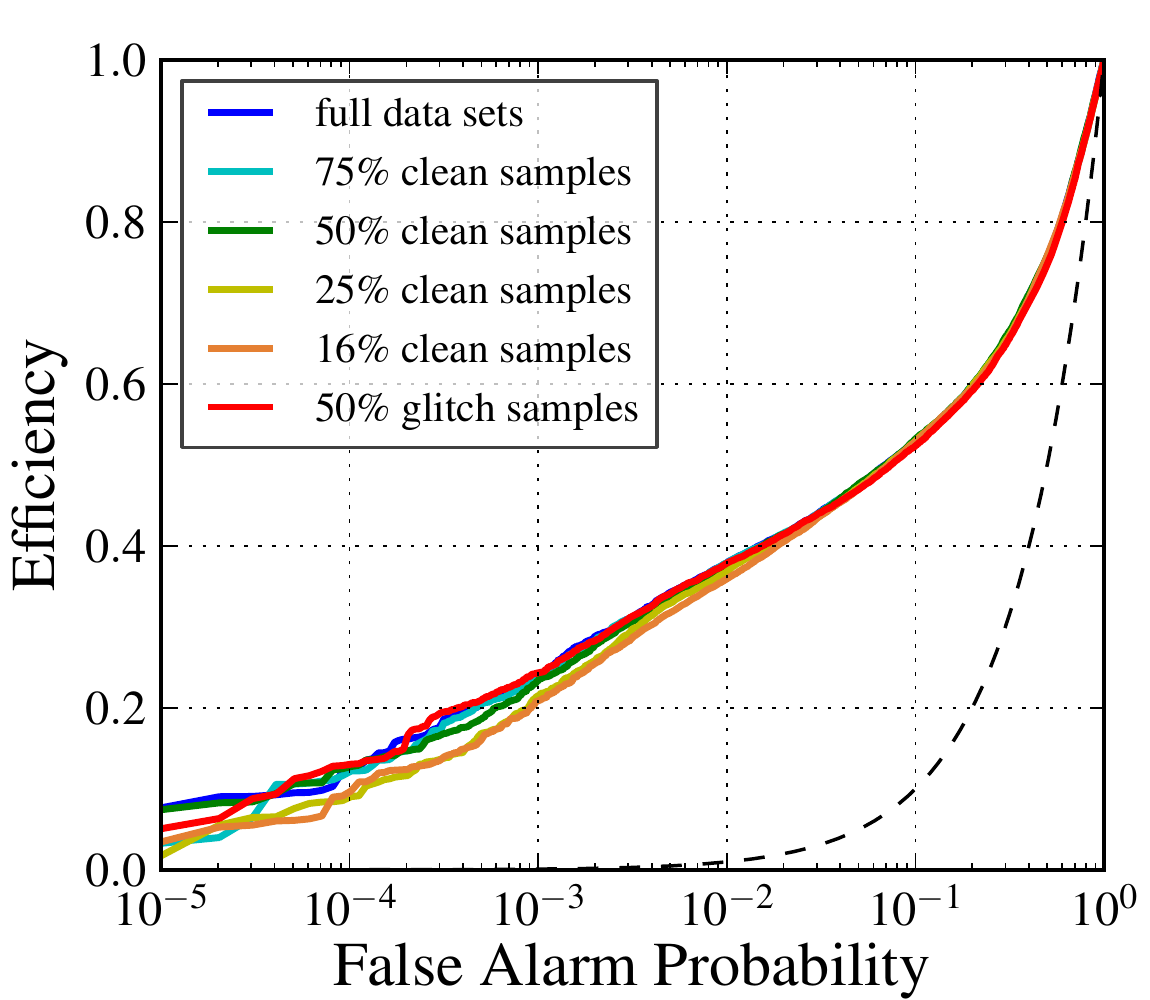}}\\
\subfloat[S6 \ac{ANN}]{\label{fig:S6_ANN_var_train_data}\includegraphics[width=0.3\textwidth]{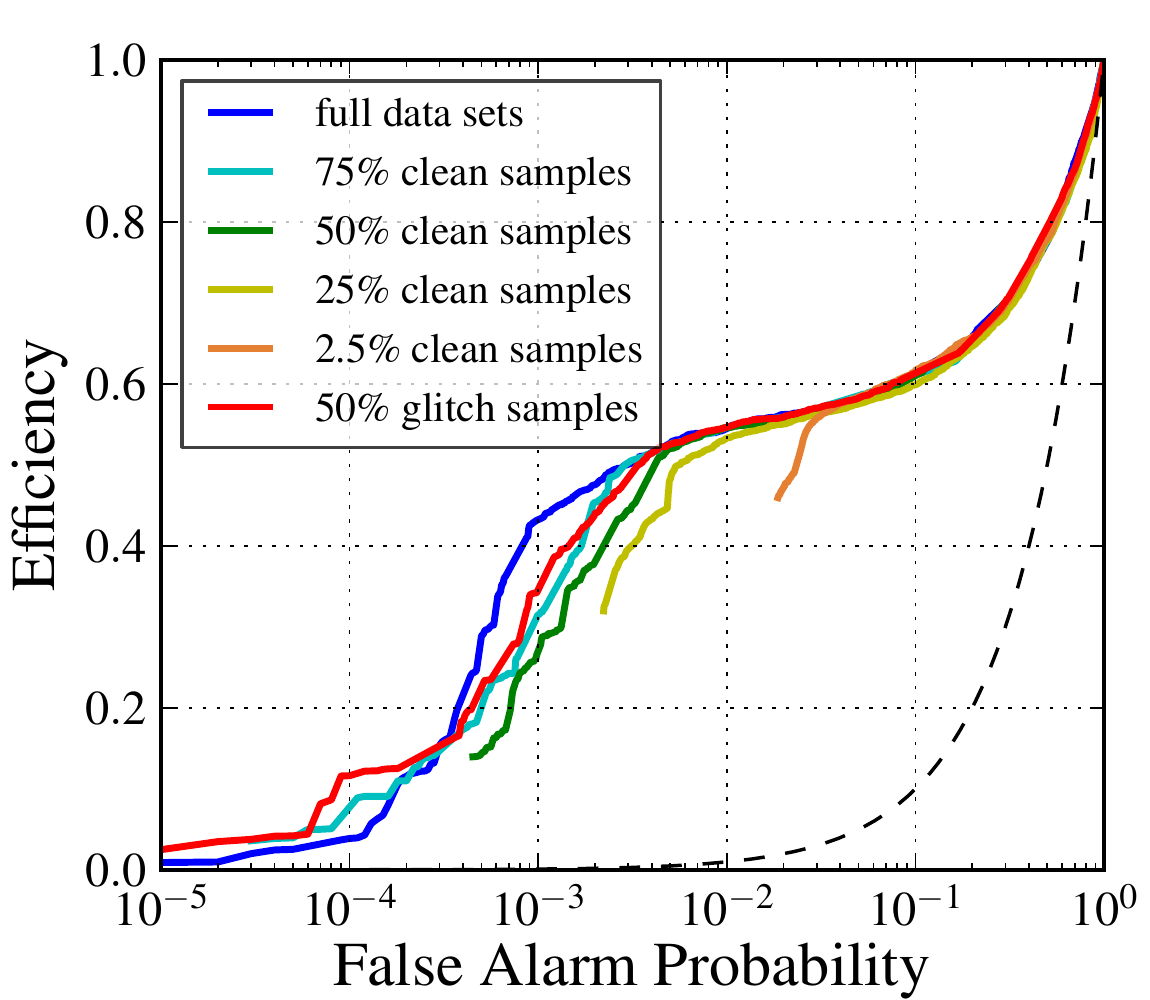}}
\subfloat[S6 \ac{SVM}]{\label{fig:S6_SVM_var_train_data}\includegraphics[width=0.3\textwidth]{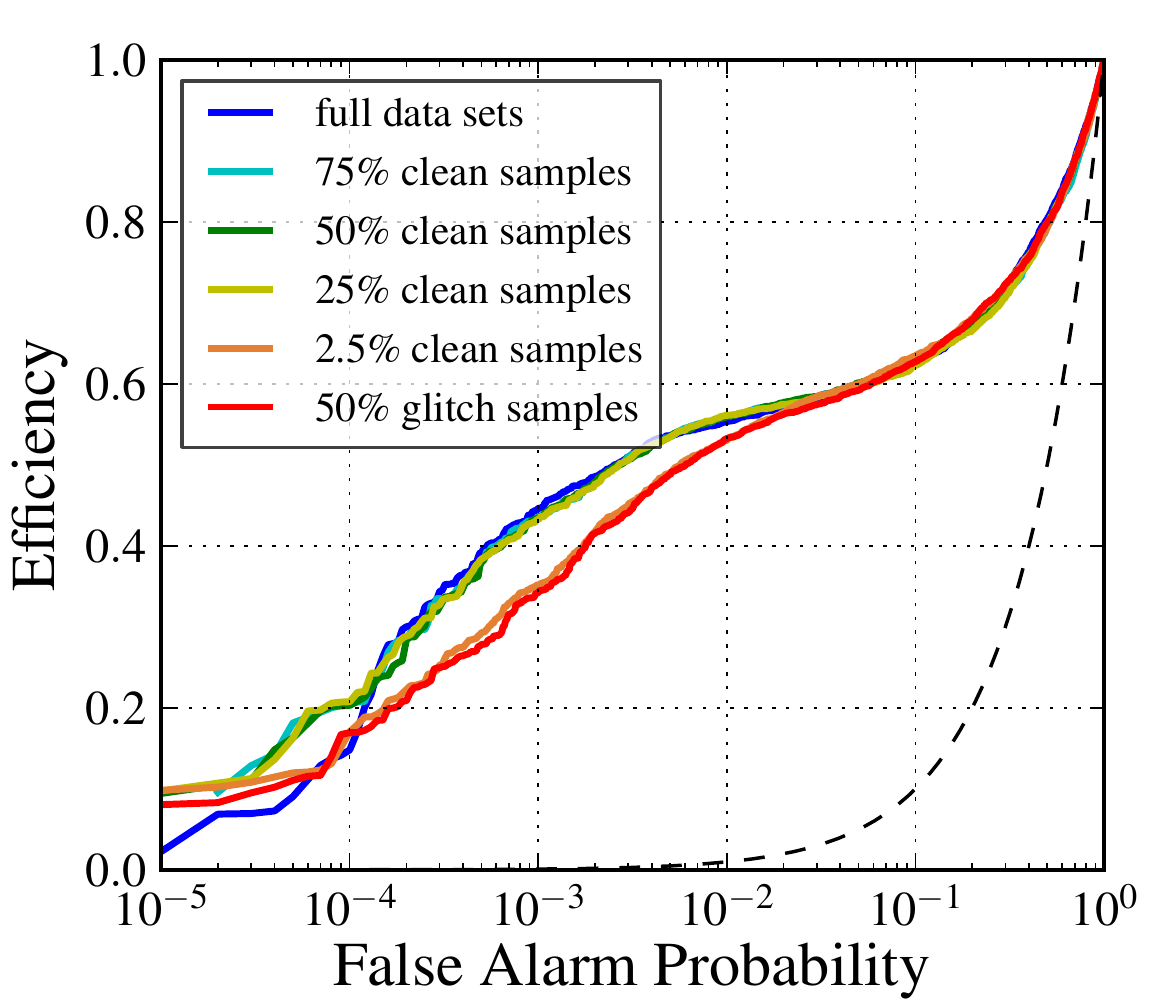}}
\subfloat[S6 \ac{RF}]{\label{fig:S6_MVSC_var_train_data}\includegraphics[width=0.3\textwidth]{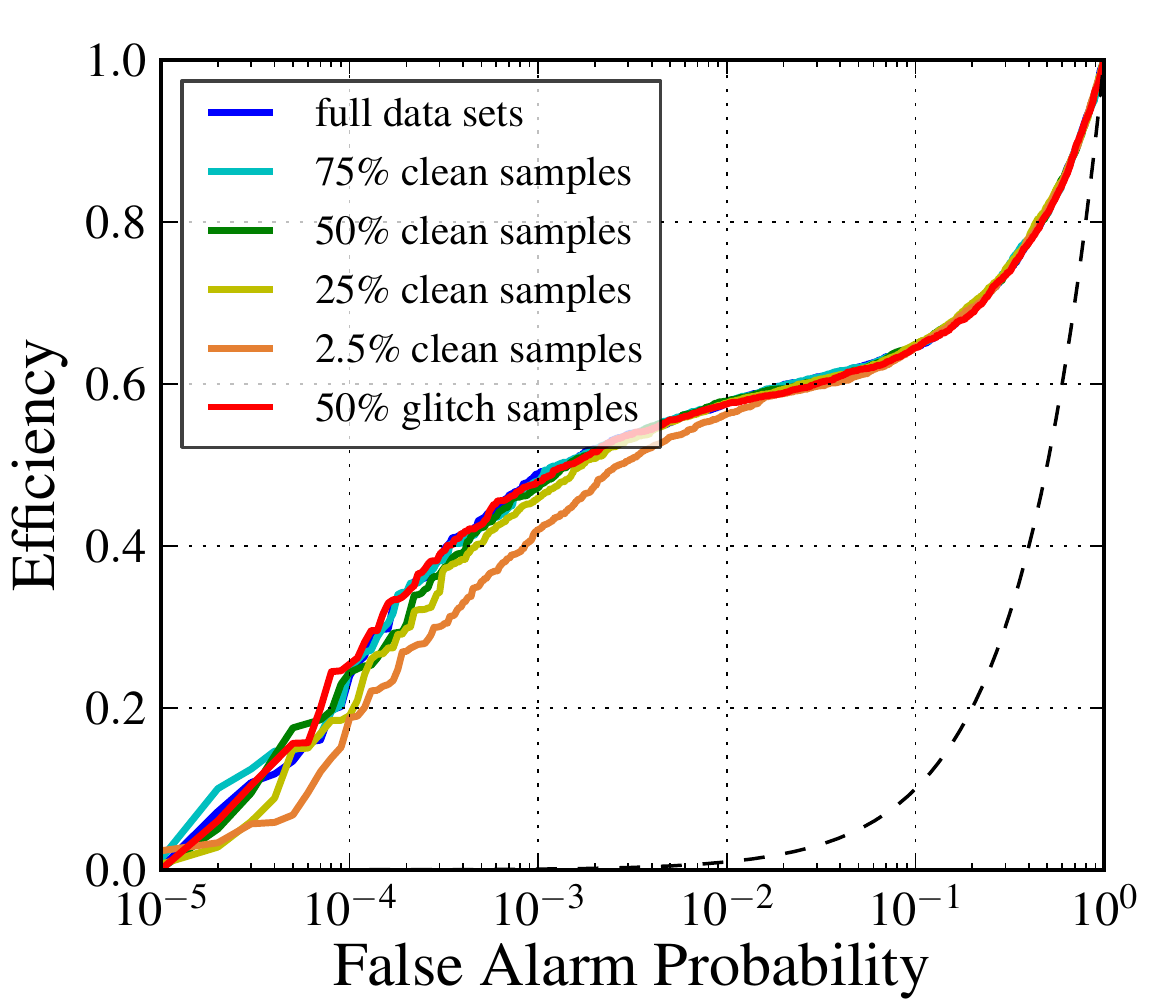}}
\caption{Varying the size of training data sets. In our sample data, the number of glitches is limited by the actual glitch rate in the \ac{LIGO} detectors and the length of the analysis time we use. However, we can construct as many clean samples as necessary because we sample the auxiliary channels at random times. In general, classifiers' performance will increase with larger training data sets, but the computational cost of large data sets will also increase. We investigate the effect of varying the size of training sets on the classifiers' performance, and observe only small changes even when we significantly reduce the number of clean samples. We also reduce the number of glitch samples, and observe that the classifiers are more sensitive to the number of glitches. This is likely due to the smaller number of glitch samples, and reducing the number of glitches may induce severe undersampling of feature space. The black dashed line represents a classifier that is based on random choice.}
\label{fig:var_train_data}
\end{figure*}

We test how the classifiers' performance is affected by the size of the clean training set in a series of runs in which we gradually reduce the number of clean samples available. Runs with 100\%, 75\%, 50\% and 25\% of the total number of clean samples available for training are supplemented by a run in which the number of clean training samples is equal to the number of glitch training samples (16\% in S4 data and 2.5\% in S6 data). In addition, we performed one run in which we reduced the number of glitch training samples by half, but kept 100\% of the clean training samples. While not completely exhaustive, we believe these runs provide us with enough information to describe the classifiers' behavior. In all of these runs, we use all available samples for evaluation, employing the round-robin procedure. \Cref{fig:var_train_data} demonstrates changes in the \ac{ROC} curves due to the variation of training sets.

\ac{RF} performance (\Cref{fig:S4_MVSC_var_train_data} and \Cref{fig:S6_MVSC_var_train_data}) is not affected by reduction of the clean training set in the explored range, with the only exception of the run over S6 data where size of the clean training set is to 2.5\% of the original. In this case, the \ac{ROC} curve shows an efficiency loss on the order of 5\% at a false alarm probability of $P_{0}=10^{-3}$. Also, cutting the glitch training set by half does not affect \ac{RF} efficiency in either S4 or S6 data. 
\ac{SVM}'s performance follows very similar trends, shown in \Cref{fig:S4_SVM_var_train_data} and \Cref{fig:S6_SVM_var_train_data}. It demonstrates robust performance against the reduction of the clean training set, suffering appreciable loss of efficiency only in the case of the smallest, 2.5\% set. Unlike \ac{RF}, \ac{SVM} seems to be more sensitive to variations in size of glitch training set. The \ac{ROC} curve for the 50\% glitch set in S6 data drops 5\%-10\% in the false alarm probability region of $P_{0}=10^{-3}$ (\Cref{fig:S6_SVM_var_train_data}). However, this does not happen in the S4 run (\Cref{fig:S6_SVM_var_train_data}). This can be explained by the fact that S4 glitch data set has five times more samples than S6 set. Even after cutting it in half, the S4 set provides better sampling than the full S6 set. 

\ac{ANN} is affected most severely by training set reduction (\Cref{fig:S4_ANN_var_train_data} and \Cref{fig:S6_ANN_var_train_data}). First, its overall performance visibly degrades with the size of the clean training set, especially in the S6 runs (\Cref{fig:S6_ANN_var_train_data}). Although, we note that the \ac{ROC} curve primarily drops near a false alarm probability of $P_{0}=10^{-3}$, it remains the same near $P_{0}=10^{-2}$ (for all but the 2.5\% set). The higher $P_0$ value is more important in practice because the  probability of false alarm of  $10^{-2}$ is still tolerable and, at the same time, the efficiency is significantly higher than at $P_{0}=10^{-3}$. This means that we are likely to operate a real-time monitor near $P_0=10^{-2}$ rather than near $10^{-3}$. Reducing the training sample introduces an artifact on \ac{ANN}'s \ac{ROC} curves, not seen on either \ac{RF} or \ac{SVM}. Here, the false alarm probability's range decreases with the size of the clean training set. This is due to the fact that with the \ac{ANN} configuration parameters used in this analysis, \ac{ANN}'s rank becomes more degenerate when less clean samples are available for training, meaning that multiple clean samples in the evaluation set are assigned exactly the same rank. This is in general undesirable, because a continuous, non-degenerate rank carries more information and can be more efficiently incorporated into gravitational-wave searches. The degeneracy issue of ANN and its possible solutions are treated in detail in \cite{Kim:2012}.

We would like to highlight the fact that in our test runs, we use data from two different detectors and during different science runs, and that we test three very different classifiers.  The common  trends that we observe are not the result of peculiarities in a specific data set or an algorithm. It is reasonable to expect that they reflect generic properties of the detectors' auxiliary data as well as the \ac{MLA} classifiers. Extrapolating this to the future applications in advanced detectors, we find it reassuring that the classifiers, when suitably configured, are able to monitor large numbers of auxiliary channels while ignoring irrelevant channels and features. Furthermore, their performance is robust against variations in the training set size. In the next sections we compare different classifiers in their bulk performance as well as in sample-by-sample predictions using the full data sets.

%%%%%%%%%%%%%%%%%%%%%%%%%%%%%%%%%%%%%%%%%%%%%%%%%%%%%%%%%%%%%%%%%%%%%%%%%%%%%%%%%%%%%
\section{Evaluating and comparing classifiers' performance}
\label{comparison}
%%%%%%%%%%%%%%%%%%%%%%%%%%%%%%%%%%%%%%%%%%%%%%%%%%%%%%%%%%%%%%%%%%%%%%%%%%%%%%%%%%%%%

The most relevant measure of any glitch detection algorithm's performance is its detection efficiency, the fraction of  identified glitches, $P_{1}$, at some probability of false alarm, $P_{0}$. The \ac{ROC} curve is the key figure of merit and can be used to assess the algorithm's efficiency and objectively compare it to other methods. It provides an estimate of the algorithm's efficiency throughout the entire range of the probability of false alarm. (The upper limit for acceptable values of probability of false alarm depends on application.) In the problem of glitch detection in gravitational-wave data, we set this value to be  $P_{0}=10^{-2}$, which corresponds to 1\% of true gravitational-wave transients falsely labeled as glitches. Another way to interpret this is that 1\% of the clean science data are removed from searches for gravitational waves. 

Our test runs, described in the previous section, demonstrate the robustness of
the \ac{MLA} classifiers against the presence of irrelevant features in the
input data. We are interested in measuring a classifiers' efficiency in the
common case where no prior information about relevance of the auxiliary
channels is assumed. For this purpose, we use the full S4 and S6 data sets,
including all channels with a wide selection of parameters. Using exactly
the same training/evaluation sets for all our classifiers allows us to assign
four ranks, ($\ANNrank$, $\SVMrank$, $\RFrank$, $\OVLrank$), one from each of
the classifiers, to every sample and compute the probability of false alarm,
$P_{0}(r_i)$ and efficiency, $P_{1}(r_i)$. While the ranks can not be compared
directly, these probabilities can. Any differences in classifiers' predictions,
in this case, are from the details and limitations of the methods themselves,
and are not from the training data.

 Glitch samples that are separated in time by less than a second are likely to be caused by the same auxiliary disturbance. Even if they are not, gravitational-wave transient candidates detected in a search are typically ``clustered'' with a time window ranging from a few hundred milliseconds to a few seconds, depending on the length of the targeted gravitational-wave signal. Clustering  implies that among all candidates within the time window, only the one with highest statistical significance will be retained. In order to avoid double counting of possibly correlated glitches and to replicate conditions similar to a real-life gravitational-wave search, we apply a clustering procedure to the glitch samples with a one second time window. In this time window, we keep the sample with the highest significance, $\rho$, of the transient in gravitational-wave channel. 

\begin{figure}[h!]
\centering
\subfloat[S4 \ac{ROC} curves]{\label{fig:S4_ROC_curves}\includegraphics[width=0.9\columnwidth]{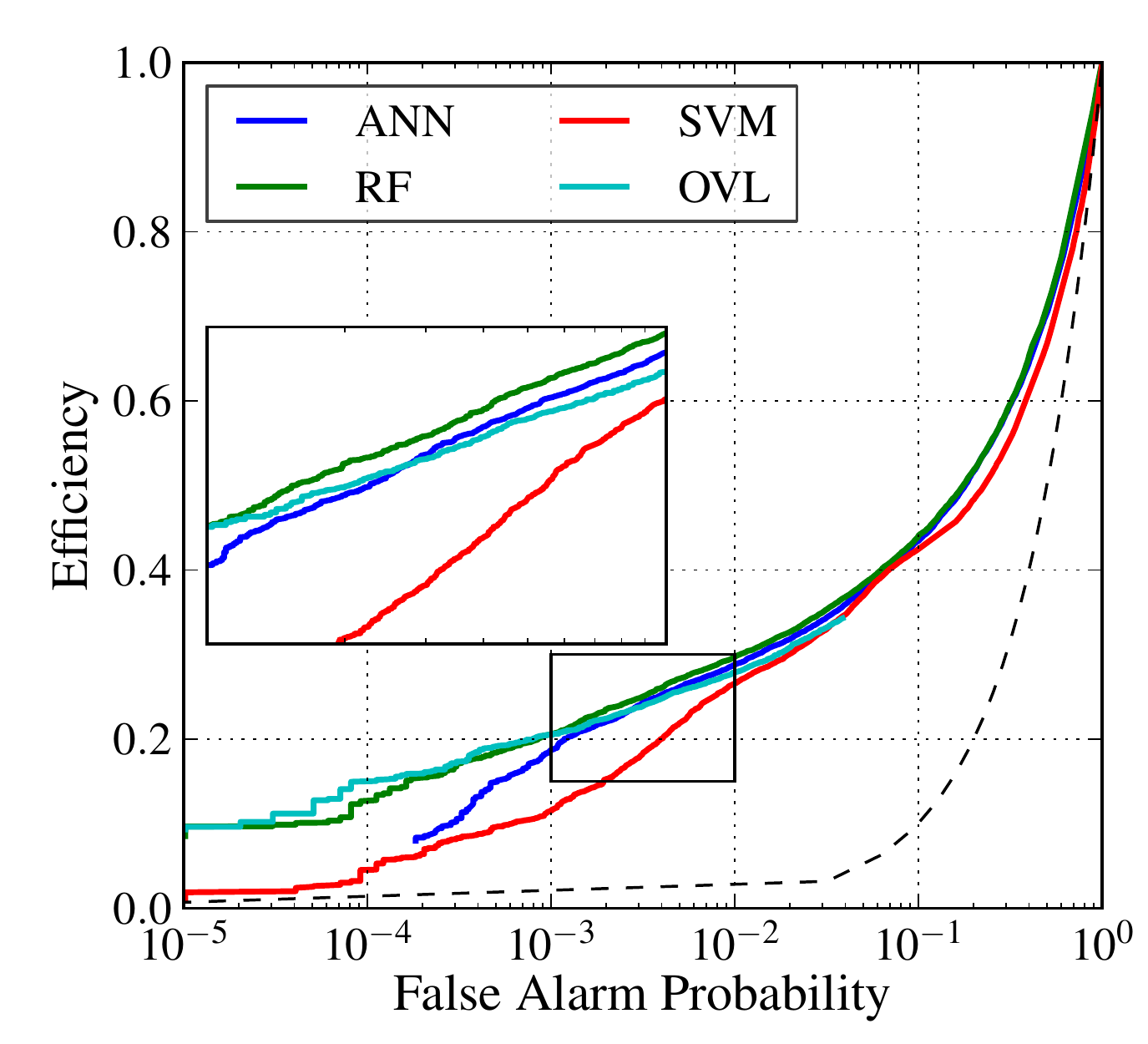}}\\
\subfloat[S6 \ac{ROC} curves]{\label{fig:S6_ROC_curves}\includegraphics[width=0.9\columnwidth]{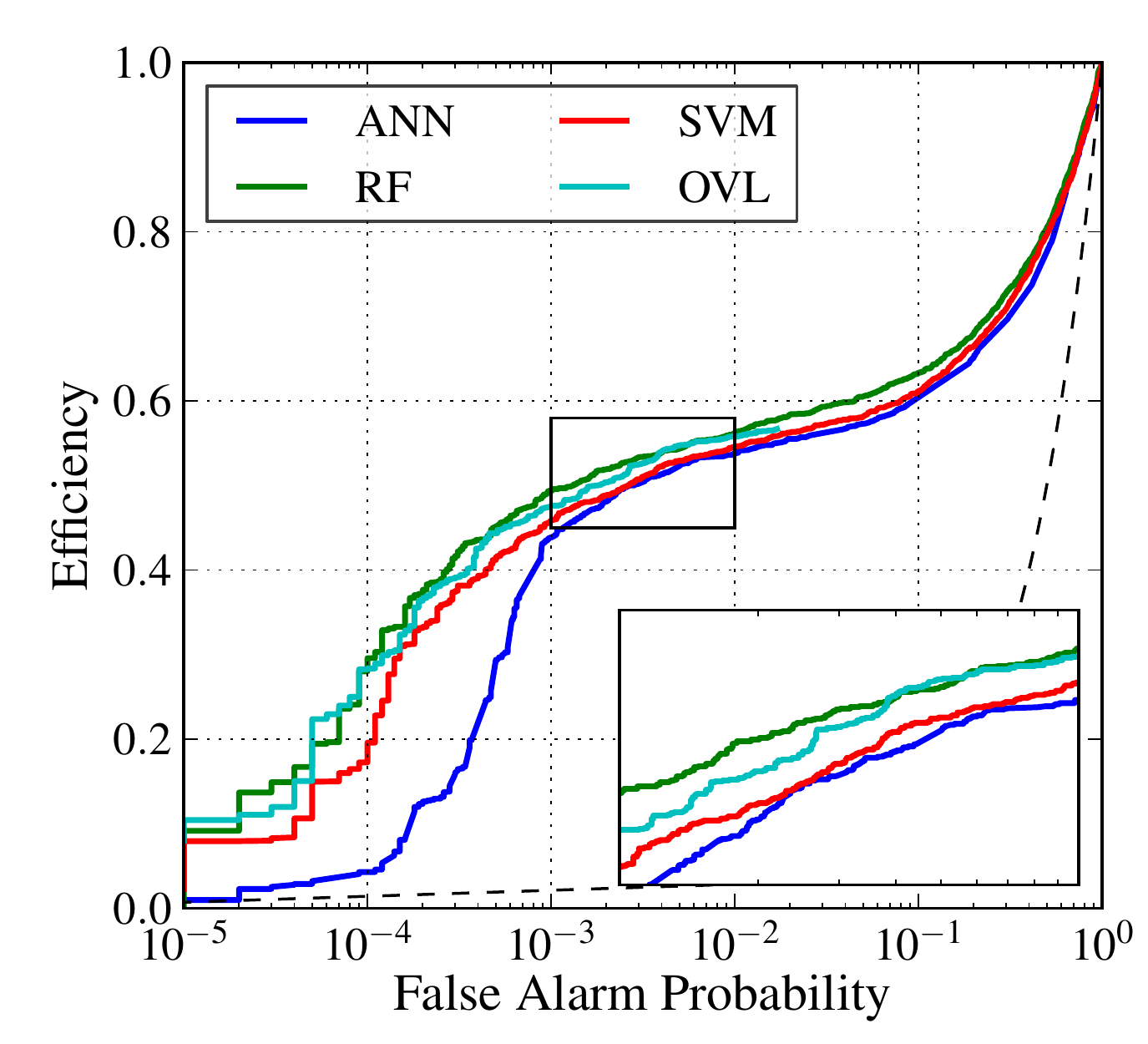}}
\caption{Comparing algorithmic performance. We directly compare the performance of \ac{RF} (green), \ac{ANN} (blue), \ac{SVM} (red), and \ac{OVL} (light blue) using the full data sets. Glitches are clustered in time. We see that all the classifiers perform similarly, particularly in S6. There is a general trend of higher performance in S6 than in S4, which we attribute to differences in the types of glitches present in the two data sets. We should also note that all the MLA classifiers achieve performance similar to our benchmark, \ac{OVL}, but \ac{RF} appears to perform marginally better for a large range of the False Alarm Probability. The dashed line corresponds to a classifier that is based on random choice.}
\label{fig:ROC_curves}
\end{figure}

The  \ac{ROC} curves  are computed after clustering. \Cref{fig:ROC_curves} shows the \ac{ROC} curves for \ac{ANN}, \ac{SVM}, \ac{RF} and \ac{OVL} for both S4 and S6 data. 

% including figures that compare ROC curves from the methods

%\begin{figure}[h!]
%	\begin{subfigure}[h]{\columnwidth}
%                \centering
%                \includegraphics[width=.9\textwidth]{Comparing_ROC_curves_S4-100ms-clustered}
%                \caption{S4 ROC curves }
%                \label{fig:S4_ROC_curves}
%        \end{subfigure}\\
%          %add desired spacing between images, e. g. ~, \quad, \qquad etc. 
%         %(or a blank line to force the subfigure onto a new line)
%		  \begin{subfigure}[h]{\columnwidth}
%                \centering
%                %place holder S4 SVM plots not generated yet, we use MVSC plots for now
%                \includegraphics[width=.9\textwidth]{Comparing_ROC_curves_S6-100ms-clustered}
%                \caption{S6 ROC curves}
%                \label{fig:S6_ROC_curves}
%         \end{subfigure}
%        \caption{Comparing algorithmic performance. We directly compare the best performance for \ac{RF} (green), \ac{ANN} (blue), \ac{SVM} (red), and \ac{OVL} (light blue) using the full data sets. We see that all the classifiers perform similarly, particularly in S6. There is a general trend of higher performance in S6 than in S4, which we attribute to differences in the types of glitches present in the two data sets. We should also note that all the MLA classifiers achieve performance similar to our benchmark, \ac{OVL}, but \ac{RF} appears to perform marginally better for a large range of the False Alarm Probability. The dashed line corresponds to a classifier that is based on random choice.}\label{fig:ROC_curves}
%\end{figure}

All our classifiers show comparable efficiencies in the most relevant range of the probability of false alarm for practical applications ($10^{-3}$ -- $10^{-2}$). Of the three \ac{MLA} classifiers, \ac{RF} achieves the best efficiency in this range, with \ac{ANN} and \ac{SVM} getting very close near $P_{0} =10^{-2}$. Relative to other classifiers, \ac{SVM} performs worse in the case of S4 data, and \ac{ANN}'s efficiency drops fast at $P \le 10^{-3}$. The most striking feature on these plots is how closely the \ac{RF} and the \ac{OVL} curves follow each other in both S4 and S6 data (\Cref{fig:S4_ROC_curves} and \Cref{fig:S6_ROC_curves} respectively). In absolute terms, the classifiers achieve significantly higher efficiency for S6 than for S4 data, 56\% verses 30\% at $P_{0} = 10^{-2}$. We also note that the clustering procedure has more affect on the \ac{ROC} curves in S4 than in S6 data. In the former case, the efficiency drops by 5 - 10\% (compare to the curves in  \Cref{fig:S4_ANN_var_train_data,fig:S4_SVM_var_train_data,fig:S4_MVSC_var_train_data}), whereas in the latter it stays practically unchanged (compare to \Cref{fig:S6_ANN_var_train_data,fig:S6_SVM_var_train_data,fig:S6_MVSC_var_train_data}). The reason for this is not clear. In the context of detector evolution, the S6 data are much more relevant for advanced detectors. At the same time, we should caution that we use just one week of data from the S6 science run and larger scale testing is required for evaluating the effect of the detector's non-stationarity.    

\begin{figure}
\centering
\subfloat[S4 glitches]{\label{fig:S4_hist_signif}\includegraphics[width=0.9\columnwidth]{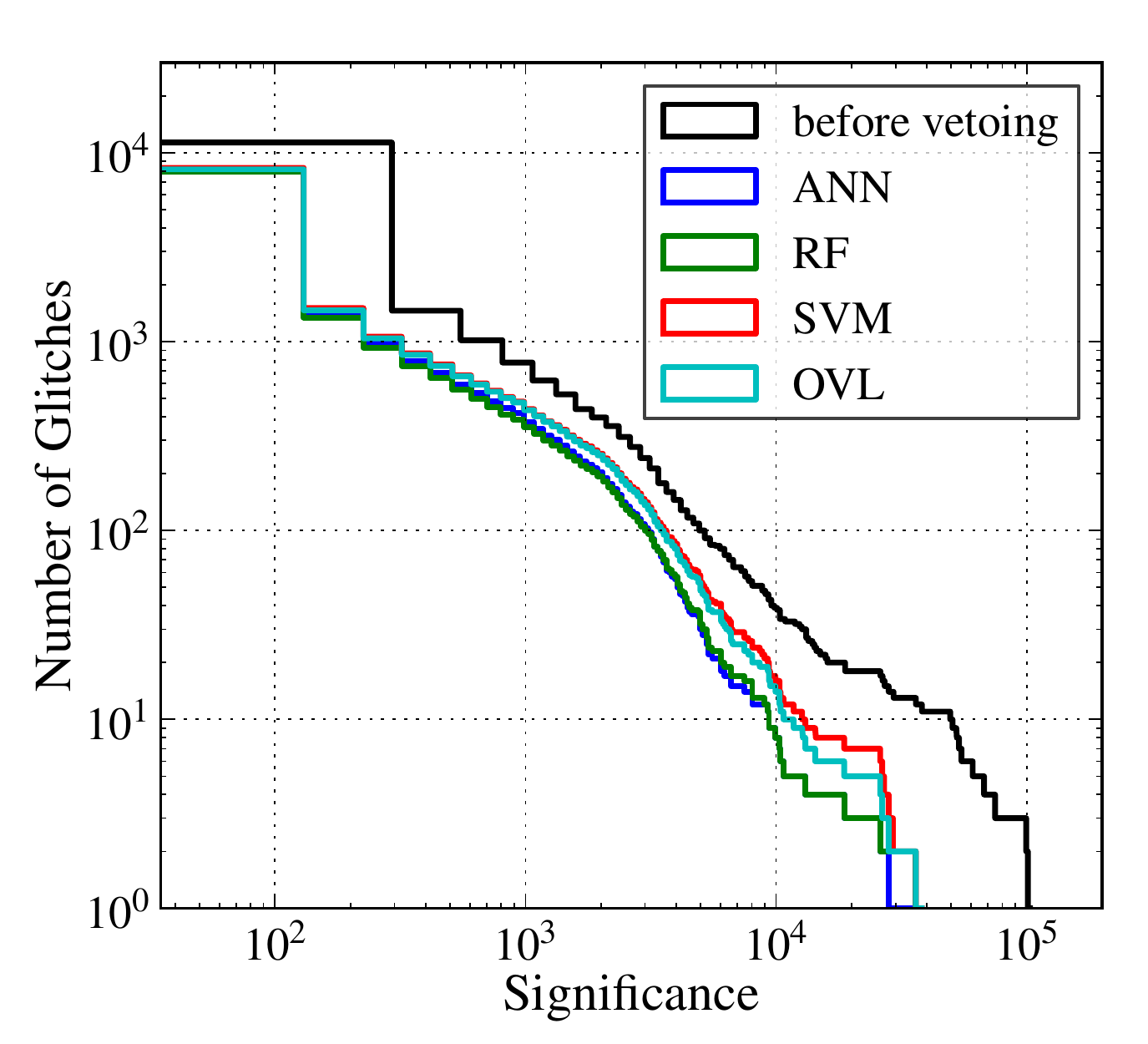}}\\
\subfloat[S6 glitches]{\label{fig:S6_hist_signif}\includegraphics[width=0.9\columnwidth]{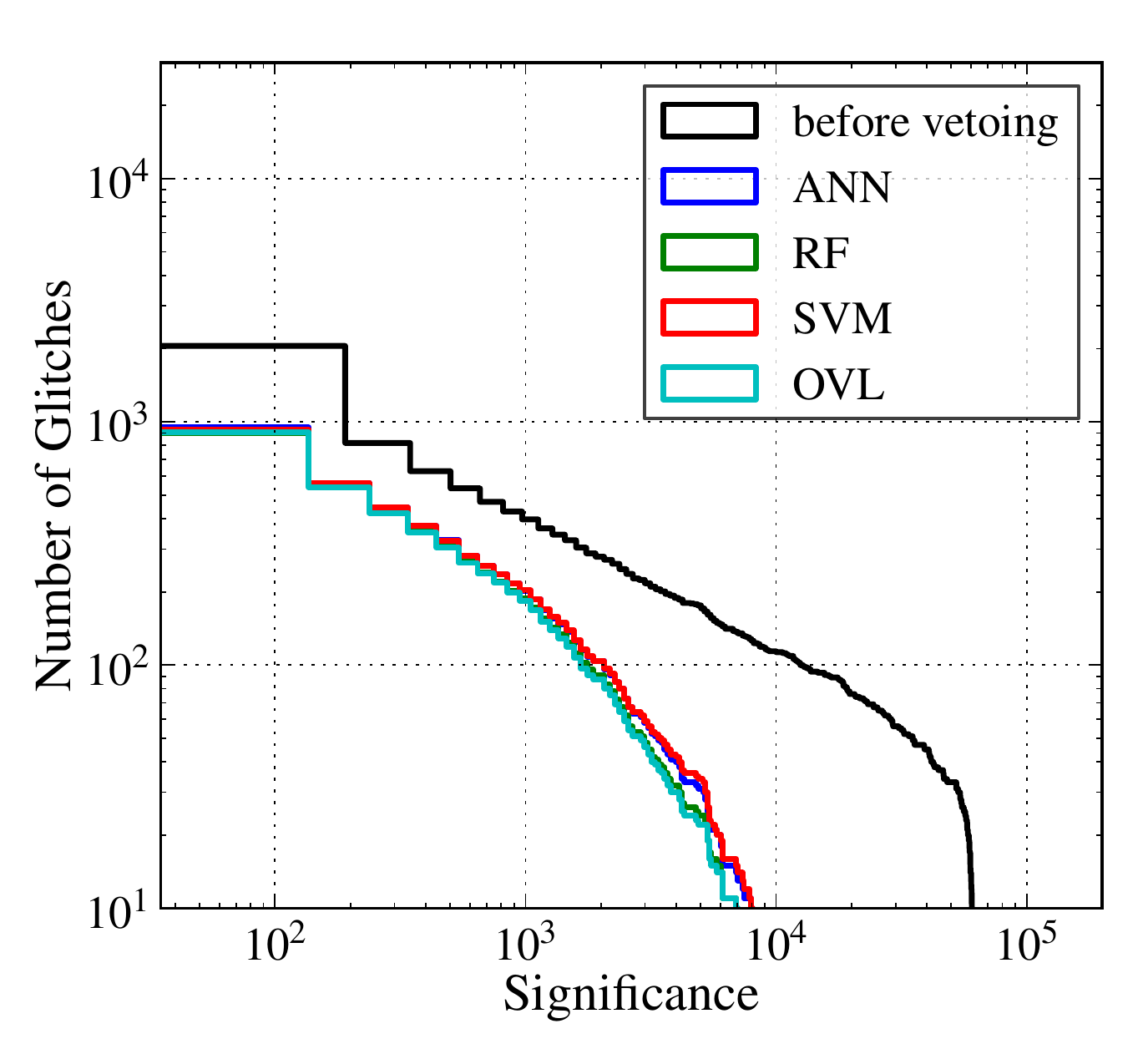}}
\caption{Comparing distribution of glitches before and after applying classifiers at 1 \% probability of false alarm. This cumulative histogram shows the number of glitches that remain with at least as high a significance in the gravitational-wave channel. We see that all our classifiers remove similar fractions of glitches at 1\% probability of false alarm. This corresponds to their similar performances in \Cref{fig:ROC_curves}, with efficiencies near 30\% and 55\% for S4 and S6 data, respectively. We also see that the classifiers tend to truncate the high-significance tails of the non-Gaussian transient distributions, particularly in S6.}
\label{fig:hist_signif}
\end{figure}

The \ac{ROC} curves characterize the bulk performance of the classifiers, but they  do not provide information about what kind of glitches are identified. To gain further insight into the distribution of glitches before and after classification, we plot cumulative histograms of the significance, $\rho$, in the gravitational-wave channel for glitches that remain after removing those detected by each of the classifiers at $P_{0} \le 10^{-2}$. We also plot a histogram of all glitches before any glitch removal. These histograms are shown in  \Cref{fig:hist_signif}. They show the effect of each classifier on the distribution of glitches in the gravitational-wave channel. In both the S4 and S6 data sets, the tail of the glitch distribution, containing samples with highest significance, is reduced. At the same time, as is clear from the plots, many glitches in the mid range of significances are also removed, contributing to overall lowering of the background for transient gravitational-wave searches. The fact that our classifiers remove low significance glitches while some of the very high significance glitches are left behind indicates that there is no strong correlation between amplitude of glitches in gravitational-wave channel and their detectability. This in turn implies that we either do not provide all necessary information for identification of these high significance glitches in the input feature  vector or the classifiers somehow do not take advantage of this information. Given the close agreement between various classifiers that we observe in the \ac{ROC} curves (\Cref{fig:ROC_curves}) and the histograms of glitch distributions (\Cref{fig:hist_signif}), the former alternative seems to be more plausible. Alternatively, our choices of the thresholds and the coincidence windows that went into the construction of the feature vectors might not be optimal. Also, heretofore unincluded features characterizing the state of the detector, which may amplify transient disturbances in the auxiliary channels and induce glitches in the gravitational-wave channel, might be crucial for identifying glitches missed in the current analysis. We leave the investigation of these possibilities to future work.   

% including figures showing before/after histograms 

Although the \ac{ROC} curves (\Cref{fig:ROC_curves}) and the histograms (\Cref{fig:hist_signif}) provide strong evidence that all classifiers detect the same glitches, they do not give a clear quantitive picture of the overlap between these methods. To see this more clearly, we define subsets of glitches based on which combination of classifiers detected them with a probability of false alarm less than $10^{-2}$. We determine overlaps between the \ac{MLA} classifiers by constructing a bit-word diagram (\Cref{fig:bit_word_mvc}). It clearly demonstrates a high degree of redundancy between the classifiers. The fraction of glitches detected by all three \ac{MLA} classifiers is 91.3\% for S6 data and 78.4\% for S4 data. For comparison, we also construct a bit-word diagram for the clean samples, shown on the same figure, which are falsely identified as glitches with probability of false alarm less than $10^{-2}$. The classifiers' predictions for clean samples are distributed almost uniformly. This suggests that our classifiers select clean samples nearly independently, or at least with a much lower level of correlation than for glitches.

% including bit-word histograms comparing MVCs
\begin{figure}[h!]
\centering
\subfloat[S4 bit-word histogram for \ac{MLA}s]{\label{fig:S4_bit_word_mvc}\includegraphics[width=0.9\columnwidth]{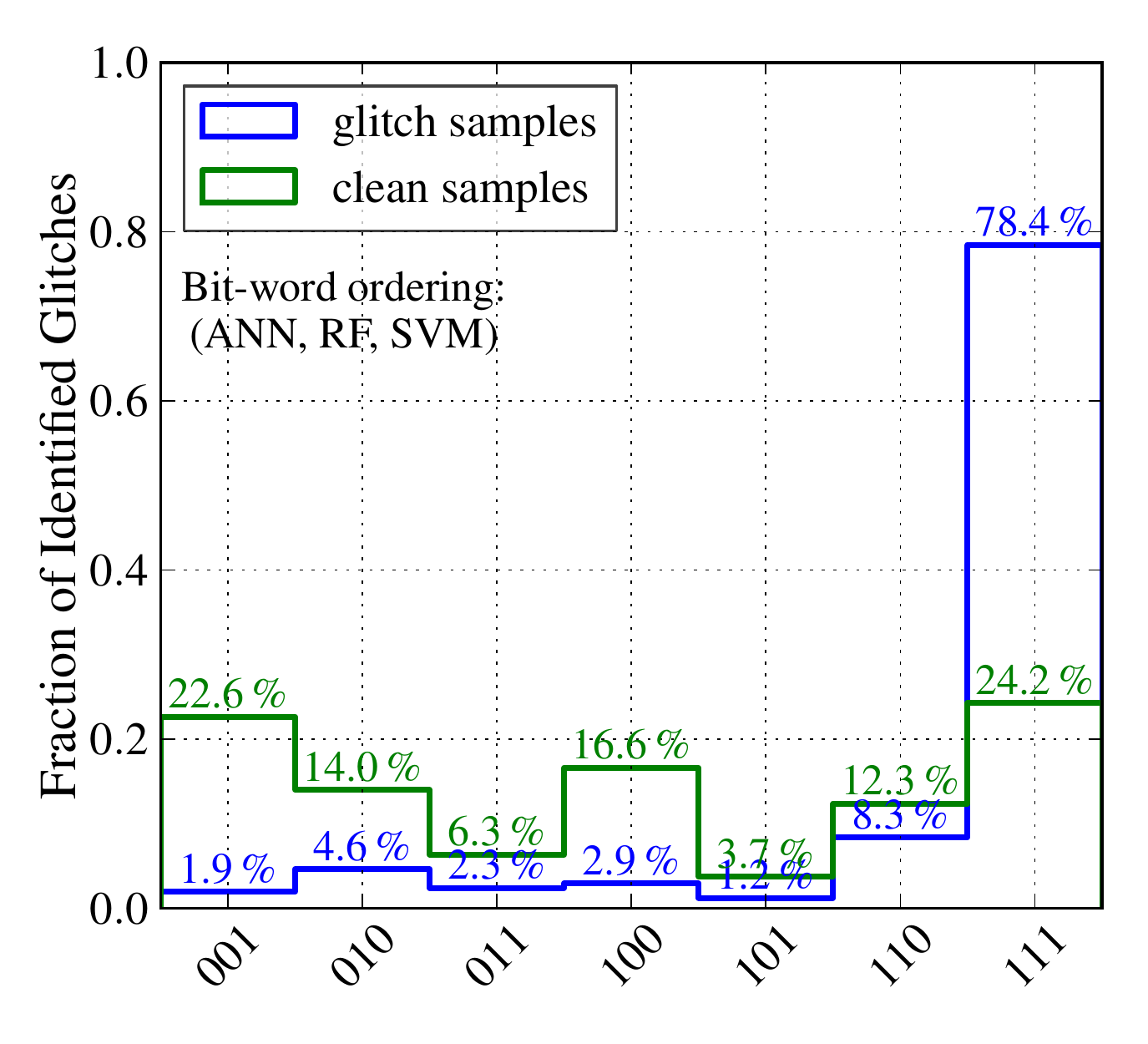}}\\
\subfloat[S6 bit-word histogram for \ac{MLA}s]{\label{fig:S6_bit_word_mvc}\includegraphics[width=0.9\columnwidth]{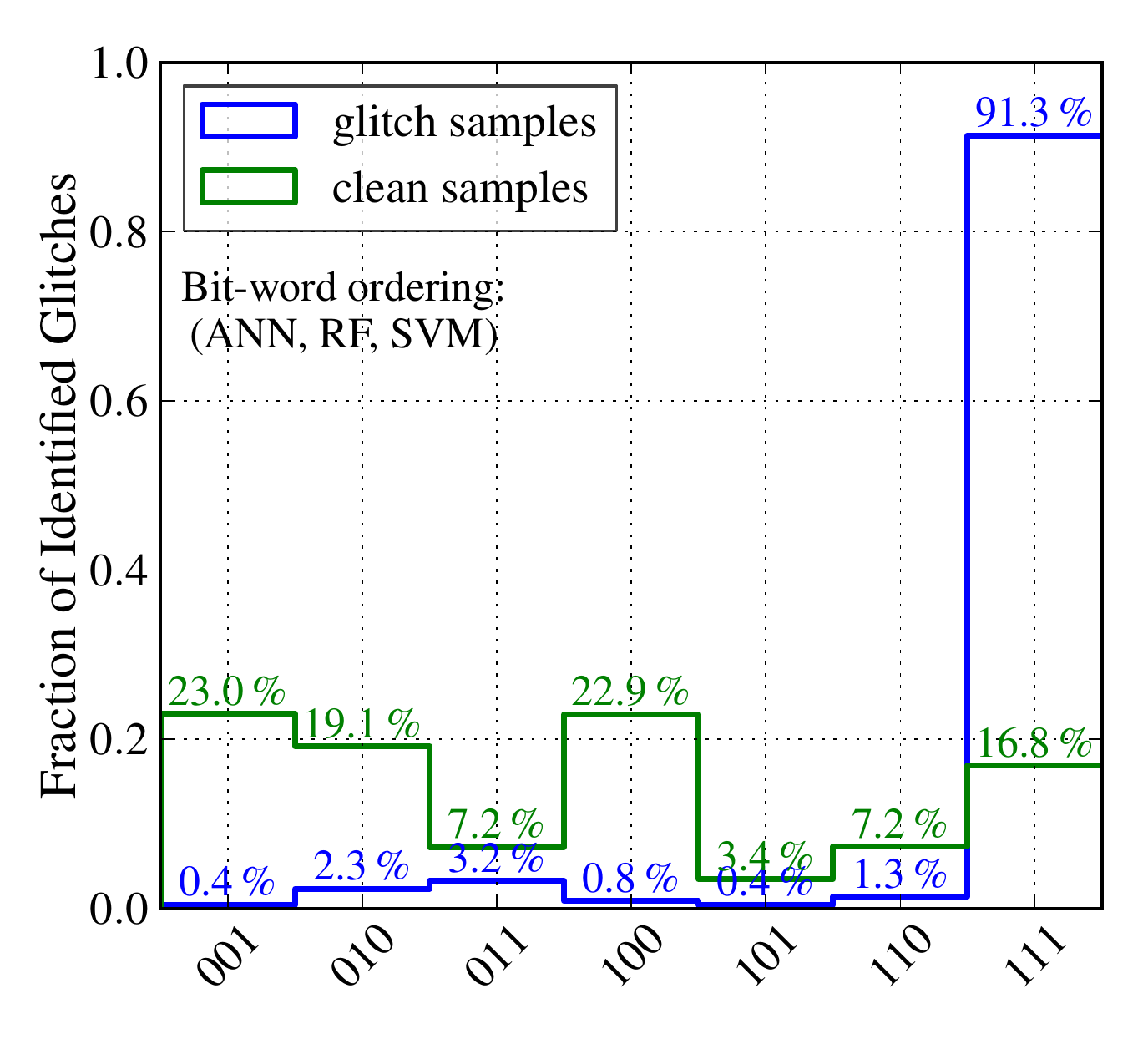}}
        \caption{Redundancy between \ac{MLA} classifiers. These histograms show
            the fractions of detected glitches identified in common by a given
            set of classifiers at 1\% probability of false alarm (blue). The
            abscissa is labeled with bit-words, which are indicators of which
            classifier(s) found that subset of glitches (eg: 011 corresponds to
            glitches that were not found by \ac{ANN}, but were found by \ac{RF}
            and \ac{SVM}). The quoted percentages represent the fractions of
            detected glitches so that 100\% represents those glitches which were
            successfully identified by at least one of the classifiers at 1\%
            false-alarm probability. The three classifiers show large overlap
            for glitch identification (bit-word $= 111$), meaning the classifiers
            are largely able to identify the same glitch events. Also shown is the
            fraction of clean samples (green) misidentified as glitches, which
            shows a comparatively flat distribution across classifier
            combinations.}\label{fig:bit_word_mvc}
\label{fig:bit_word_mvc}
\end{figure}

\begin{figure}[h!]
\centering \subfloat[S4 bit-word histogram for \MLAmax and
\ac{OVL}]{\label{fig:S4_bit_word_ovl}\includegraphics[width=0.9\columnwidth]{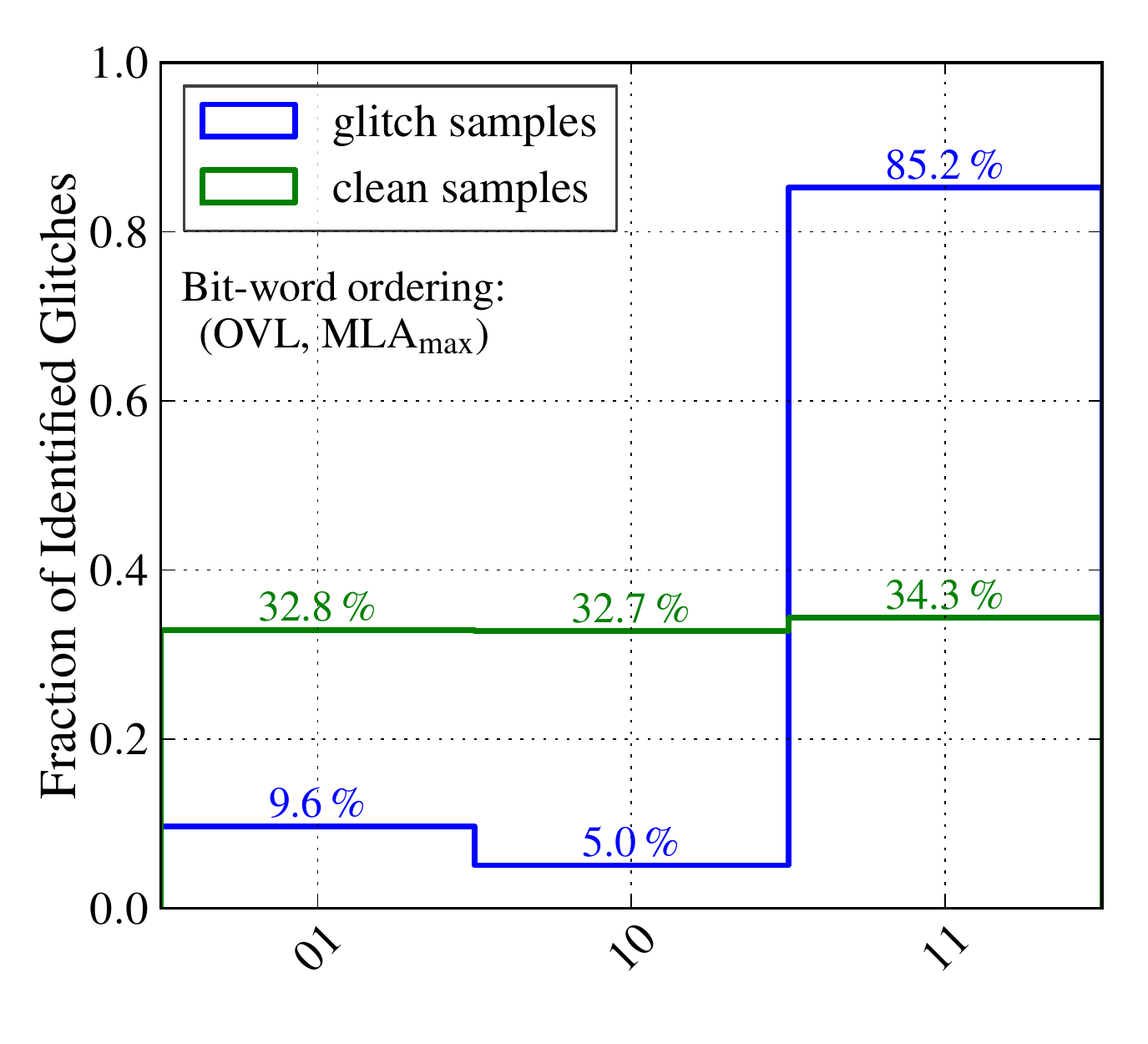}}\\
\subfloat[S6 bit-word histogram for \MLAmax and
\ac{OVL}]{\label{fig:S6_bit_word_ovl}\includegraphics[width=0.9\columnwidth]{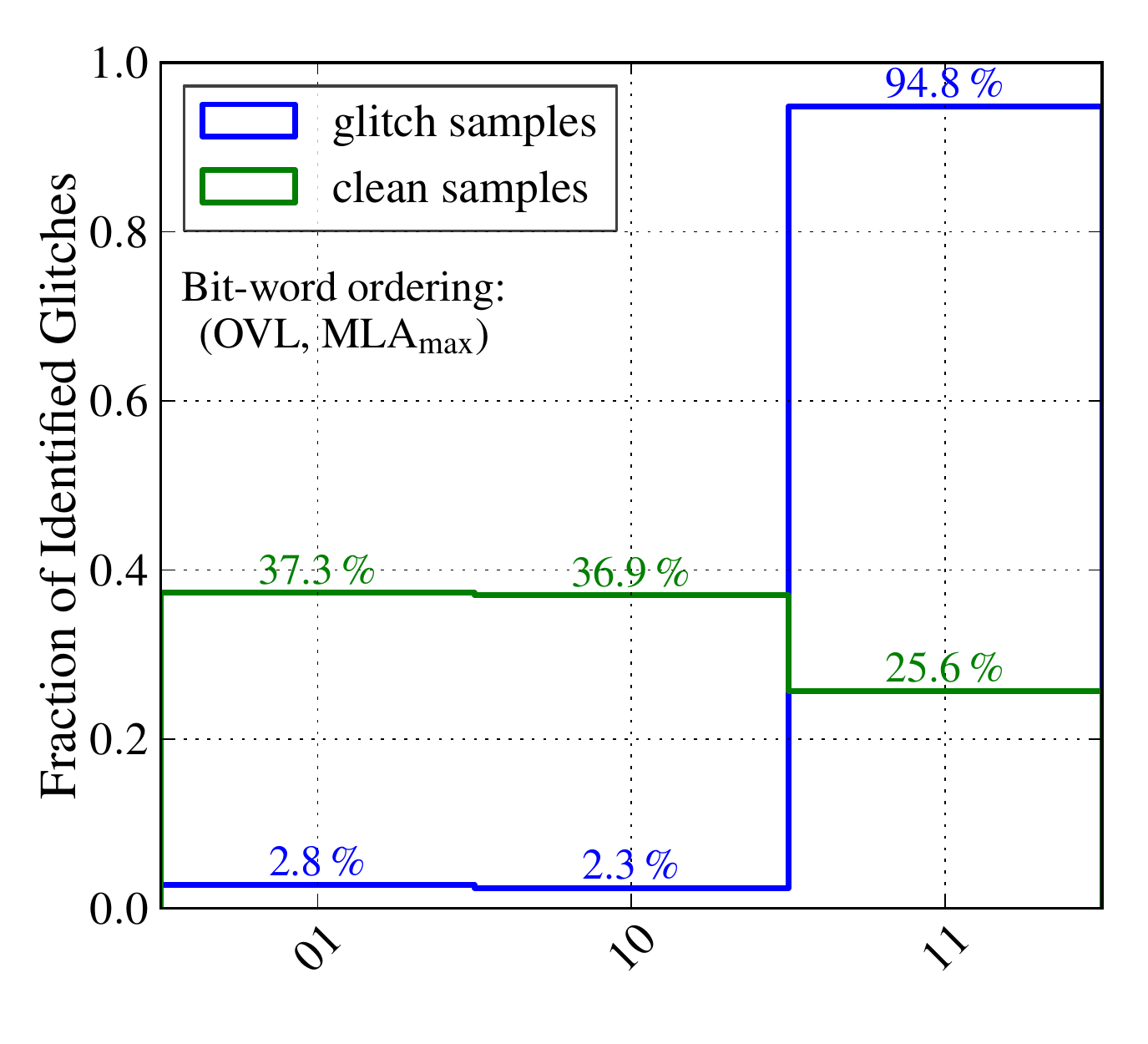}}
\caption{Redundancy between \MLAmax\ and OVL. This figure is similar to
	\Cref{fig:bit_word_mvc}, except these histograms only compare the results
	of combining the MLA classifiers into a single unified classifier (\MLAmax)
	and \ac{OVL}. Even though \ac{OVL} only considers pairwise correlations
	between auxiliary channels and the gravitational-wave channel, we see that
	it identifies the same glitches as \MLAmax. This suggests that the glitches
	identified by the MLA classifiers are effectively characterized by pairwise
	correlations between a single auxiliary channel and the gravitational-wave
	channel, and that considering multi-channel correlations does not add much.
	We also see that these classifiers are highly correlated on their selection
	of glitches (blue), but less correlated across the set of misidentified
	clean samples (green).}
\label{fig:bit_word_ovl}
\end{figure}

Next, we compare the \ac{MLA} classifiers to \ac{OVL}. In order to reduce the number of possible pairings, we combine the \ac{MLA} classifiers following the maximum-likelihood-ratio algorithm described in more detail in the next Section~\ref{combining}. In short, this algorithm picks the most statistically significant prediction out of the three \ac{MLA} classifiers for each event. We denote the combined classifier as \MLAmax. As in the previous case, we construct the bit-word diagram for both glitch and clean samples detected with the probability of false alarm less than $10^{-2}$ (\Cref{fig:bit_word_ovl}). The redundancy is even stronger. The fraction of glitches detected by \MLAmax and \ac{OVL} is 94.8\% for S6 data and  85.2\% for S4 data. The full bit-word histograms show the same behavior and we omit them here.

% including bit-word histograms comparing combined MVC with OVL

\section{Methods for combining classifiers }
\label{combining}

On a fundamental level, the \ac{MLA} classifiers search for a one parameter family of decision surfaces in the feature space, $\auxvec \in \Vdata$, by optimizing a detection criterion. The parameter labeling the decision surfaces can be mapped into a continues rank, $\MLArank(\auxvec) \in [0,1]$. The rank reflects the odds for a sample, $\auxvec$, to correspond to a glitch in the gravitational-wave channel. As we discuss in Section~\ref{detection_problem} and Appendix~\ref{appendix:fom}, theoretically, if the classifiers use consistent optimization criteria, they should arrive at the same optimal decision surfaces and make completely redundant predictions. In other words, their ranks would be functionally dependent. In practice, however, different classifiers often lead to different results, primarily due to the limitations in the number of samples in the training sets and/or computing resources. For instance, different classifiers may be more or less sensitive to different types of glitches. In this case, one should be able to detect a larger set of glitches by combining their output. Furthermore, the classifiers may be strongly correlated in the ranks they assign to glitch samples, but only weakly correlated when classifying clean samples. Again, by combining the output of different classifiers, we may be able to extract information about these correlations and improve the total efficiency of our analysis.

This last case appears to be applicable to our data set. From Section~\ref{comparison}, we see that at a probability of false alarm of 1\%, all classifiers remove nearly identical sets of glitches (to within 10\% for the S6 data). However, the classifiers agree to a significantly lesser extent on the clean samples they remove (\Cref{fig:bit_word_mvc}). This suggests that the correlations between the classifiers' predictions are different for glitches and clean samples, and combining the classifiers' output could possibly lead to an improved analysis.

The general problem of combining the results from multiple, partially redundant analysis methods has been addressed in the context of gravitational-wave searches in~\cite{Biswas2012b}. Treating the output of the classifiers, namely their ranks, as new data samples, one arrives at the optimal combined ranking given by the joint likelihood ratio:

\begin{equation}
\label{joint_lr}
\Lambda_{\mathrm{joint}}(\vecrank) = \frac{p(\vecrank \given 1)}{p(\vecrank \given 0)}\,,
\end{equation}  

\noindent where $\vecrank \equiv (\ANNrank, \SVMrank, \RFrank)$ is the vector of the \ac{MLA} ranks assigned to a sample, $\auxvec$, and $p(\vecrank \given 1)$  and $p(\vecrank \given 0)$ are the probability density functions for the rank vector in the case of glitch and clean samples, respectively. We should point out that we can modify this ranking by multiplying by the ratio of prior probabilities ($p(1)/p(0)$) to match the rankings for individual classifiers without affecting the ordering assigned to samples. Typically, these conditional probability distributions are not known and computing the joint likelihood ratio from first principles is not possible. One has to develop a suitable approximation. We try several different approximations when combining algorithms.

Our first approximation, and perhaps the simplest, estimates the likelihood ratio for each classifier separately and assigns the maximum to the sample. This method should be valid in the two limits: extremely strong correlations and extremely weak correlations between the classifiers. It was first suggested and applied in the context of combining results of multiple gravitational-wave searches in~\cite{Biswas2012b}. We estimate the individual likelihood ratios in two ways: 1) as the ratio of \ac{cdf} and 2) as the ratio of kernel density estimates for the \ac{pdf}. Though a proper estimate should involve the \ac{pdf}s, the advantage of using \ac{cdf}s is that we already calculate them when evaluating the efficiency and probability of false alarm for each classifier. They should approximate the  ratio of \ac{pdf}s reasonably well in the tail of the distributions, when the probability of false alarm is low. This assumes that \ac{pdf}s are either slowly varying or simple (e.g. power law or exponential) decaying functions of the rank. 
%This assumes that the pdfs are slowly varying so that we can approximate the integrals defining the cdf's as a simple product ($\mathrm{Eff}(r) = \int_{r}^1 p(\rho|g)\,\mathrm{d}\rho \approx p(r|g) * (1-r)$), and therefore the ratio of Eff and FAP will approximate the ratio of the pdfs. 
However, at large values of the probability of false alarm or in the case when the probability distributions exhibit complicated functional dependence on the rank, our approximation may break down and we will have to resort to the more fundamental ratio of the \ac{pdf}s. Explicitly, we estimate the joint likelihood ratio using 

\begin{equation}
\label{maxcdflr}
L_{1}(\vecrank)\ \equiv \max_{r_j} \left\{\frac{\int_{r_j}^{1} p( r_j' \given 1) \diff r_j'}{\int_{r_j}^{1} p( r_j' \given 0) \diff r_j'}\right\} = \max_{r_j}\frac{P_{1}(r_j)}{P_0(r_j)}\,.
\end{equation}

\noindent We refer to this method as \MLAmax when introducing it in the context of \Cref{fig:bit_word_ovl}.

We also construct smooth one-dimensional \ac{pdf}s for clean and glitch samples from their ranks using Gaussian kernel density estimation~\cite{wand1995kernel}. These estimates were built using a constant bandwidth equal to 0.05 in the rank space, which ranges from 0 to 1. Based on this, we define the approximate combined rankings: 

\begin{equation}
\label{maxpdflr}
L_{2}(\vecrank) \equiv \max_{r_j} \left\{\frac{p( r_j \given 1)}{p( r_j \given 0)}\right\}\,.
\end{equation}

It is by no means true that we can always approximate the multi-dimensional likelihood ratio (\ref{joint_lr}) with the maximum over a set of one-dimensional likelihood ratios. If we can better model the multi-dimensional probability distributions, we should be able to extract more information. To this end, we also implement a slightly more complicated combining algorithm. We observe that the algorithms are highly correlated on which glitches they remove, and less correlated on the clean samples (see \Cref{fig:bit_word_mvc}). We therefore approximate $p(\vecrank \given 1) \approx \max_{r_j}\{p(r_j \given 1)\}$ and $p(\vecrank \given 0) \approx \prod_j p(r_j \given 0)$, which assumes that the algorithms are completely uncorrelated for the clean samples. $\Lambda_{\mathrm{joint}}$ is then approximated by

\begin{equation}
\label{randomcleanlr}
L_{3}(\vecrank) \equiv \frac{\max_{r_j} \left\{p(r_j \given 1)\right\}}{\prod_{i} p(r_i\given 0)}\,.
\end{equation} 

\noindent Again, we compute the individual \ac{pdf}s using Gaussian kernel density estimation.

More subtle, but still useful, correlations between the ranks assigned by different classifiers can not be accounted for by these simple analytical approximations. Estimating the multi-dimensional probability distributions is a difficult task and under-sampling quickly becomes the dominant source of error when expanding to higher than two dimensions. Rather than developing a complicated analytic model, we can use one of the \ac{MLA} classifiers to compute the combined rank. We use \ac{RF} to attempt to combine the ranks from each classifier and construct an estimate of the full (three-dimensional) joint likelihood ratio.

\begin{figure}
	\centering
	\includegraphics[width=.9\columnwidth]{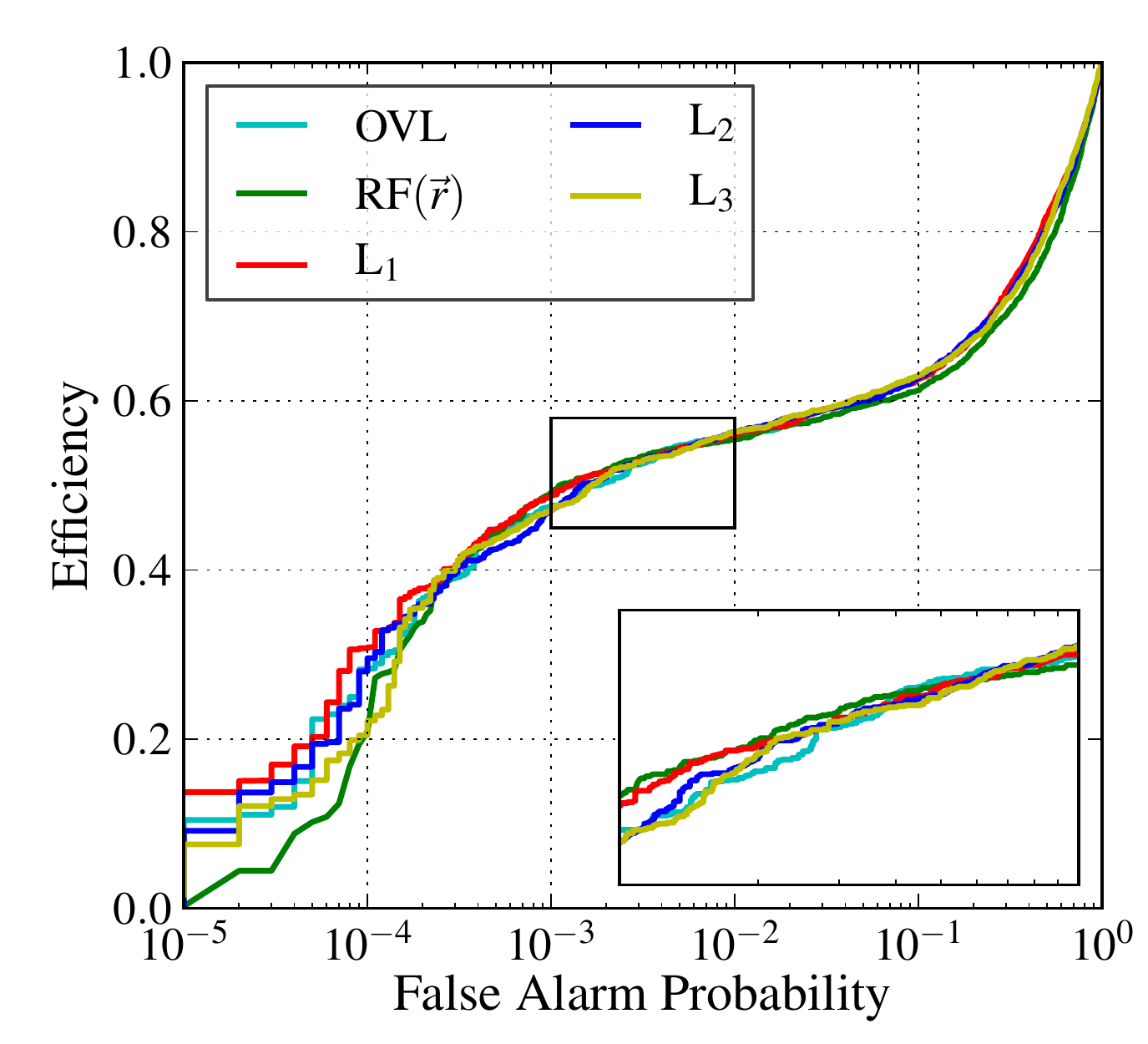}
	\caption{Comparison of different combining algorithms using S6 data. This figure compares the performance of our various schemes for combining the output of the three MLA classifiers. We note that all four algorithms,  $L_1$ (\Cref{maxcdflr}),  $L_2$ (\Cref{maxpdflr}), $L_3$ (\Cref{randomcleanlr}), and using \ac{RF} to classify events based on the MLA output vector $\vec{r}$, agree to a remarkable degree. The fact that our simple analytic algorithms perform just as well as the \ac{RF} suggests that there are not many subtle correlations between the classifiers' output. The MLA combining algorithms do not perform much better than \ac{OVL}. Comparing these curves with \Cref{fig:ROC_curves} shows that the combined performance does not exceed the individual classifier's performances. This suggest that the individual MLA classifiers each extract almost all of the useful information from our feature vectors, and that they identify the same types of glitches. These conclusions are further supported by \Cref{fig:bit_word_mvc}.}
	\label{fig:S6_combining_comparison}
\end{figure}

We compare the methods for combining the classifiers by computing \ac{ROC} curves, which are shown in~\Cref{fig:S6_combining_comparison}. We reproduce only the S6 curves because the S4 data shows the same trends. %We computed the \ac{ROC} curves for our approximations to the joint likelihood ratio, $L_{i}$'s, and found sbut did not pursue 

%\begin{figure}[h!]

All combined methods result in very similar \ac{ROC} curves and, when compared to the \ac{OVL} curve, they do not seem to improve the overall performance by more than a few percent. These combined results lead us to conclude that the individual classifiers have already reached nearly optimal performance for the given input data, and that their combination, while increasing their robustness, can not improve the overall efficiency. Basically, all the useful information has been extracted already.      

Although it is not immediately apparent, these combining schemes do add robustness to our identification of glitches. The combining algorithms are able to ignore underperforming classifiers and reject noisy input fairly well, and we see that they tend to select the best performance from the individual classifiers. By comparing \Cref{fig:S6_combining_comparison} with \Cref{fig:ROC_curves}, we see that the combining algorithms follow the best ROC curve from figure \Cref{fig:ROC_curves}, even when individual classifiers are not performing equitably. This is most evident at extremely low probabilities of false alarm. This robustness is important because it can protect a combined glitch identification algorithm from bugs in a single classifier. In this way, the combining algorithm essentially acts as an automatic cross-reference between individual MLA classifiers.

%We should note that our estimations suffer when classifiers show sudden changes in efficiency with FAP. Essentially, the problem is one of under-sampling, and our smoothing algorithm becomes susceptible to statistical fluctuations and can produce artifically high likelihood ratio estimates. This is particularly problematic for $p_i(r_i|\,\text{clean})$, which is in the denominator of the our estimate of $\Lambda_i$ and therefore can produce large erroneous estimations. This means that our algorithm can suffer from the addition of additional classifiers if they are particularly noisy or their $\Lambda_i$ are poorly constructed. The cumulative distribution functions naturally smooth out these fluctuations and therefore are more robust against this type of error (evident in the ROC plots.)

%%%%%%%%%%%%%%%%%%%%%%%%%%%%%%%%%%%%%%%%%%%%%%%%%%%%%%%%%%%%%%%
\section{Conclusion}
\label{sec:conclusion}
%%%%%%%%%%%%%%%%%%%%%%%%%%%%%%%%%%%%%%%%%%%%%%%%%%%%%%%%%%%%%%%

In this study, we apply various machine learning algorithms to the problem of identifying transient noise artifacts (glitches) in gravitational-wave data from \ac{LIGO} detectors. Our main goal is to establish the feasibility of using \ac{MLA}s for robust detection of instrumental and environmental glitches based on information from auxiliary detector channels. We consider several \ac{MLA} classifiers: the Artificial Neural Networks, the Support Vector Machine, and the Random Forest. We formulate the general detection problem in the context of glitch identification using auxiliary channel information and show that, theoretically, all classifiers lead to the same optimal solution. In a real-life application, our classifiers have to monitor a large number of channels. Even after data reduction, the dimensionality of our feature space could be as high as 1250,  making classification a truly challenging task. We test classifiers using data sets from S4 and S6 LIGO scientific runs. We use standard \ac{ROC} curves as the main figure of merit to evaluate and  compare the classifiers' performances.

Our tests show that the classifiers can handle extraneous features efficiently without affecting their performance. Likewise, we find that the classifiers are generally robust against changes in the size of the training set. The most important result of our investigation is the confirmation that the \ac{MLA} classifiers can be used to monitor a large number of auxiliary channels, many of which might be irrelevant or redundant, without a loss of efficiency. These classifiers can be used to develop a real-time monitoring and detector characterization tool.

% We apply a reduction algorithm to the auxiliary channel data from S4 and S6 science runs of LIGO detectors, recording only statistically significant transients in these channel coincident within a narrow time window with a glitch in gravitational-wave channel. From these we build the feature vector based on which the classifiers classify the unknown events. Depending on the number of features taken from each auxiliary channel the dimensionality of the resulting data space can be as high as 1250. Potentially many of the dimensions might be either redundant or irrelevant for glitch identification. Under such conditions, robust classification of the events in such high dimensional feature space is a highly non-trivial task. 

%First, we investigate how algorithms' performances are affected by the variations in the input data. We find that algorithms efficiently handle unimportant dimensions and their performance is unaffected whether they are included or excluded from the input data. Likewise, we find that algorithms are robust against the changes in the sizes of the training sets unless they become too small. The most important result of these test is confirmation that \ac{MLA} can be used to monitor without loss of efficiency large number of the auxiliary channels of which many might be irrelevant. These can be used for development a real-time monitoring and detector characterization tool. 

After establishing the robustness of the classifiers against changes in the
input data and the presence of nuisance parameters, we evaluate the algorithms'
performance in terms of the \ac{ROC} curve and carry out detailed comparison
between the classifiers. This includes their impact on the overall distribution
of glitches in the gravitational-wave channel and the redundancy of their
predictions. We find that at a false alarm probability of 1\%, all classifiers
demonstrate comparable performance and achieve 30\% and 56\% efficiency at
identifying single-detector glitches above our nominal threshold when tested on
the S4 and S6 data respectively. While not superb, this is a step toward the
ultimate goal of being able to remove all artifacts from the data. 

One advantage of the \ac{MLA} classifiers is that they provide a continues
rank, $\MLArank \in [0,1]$, rather than a binary flag. This statistic can be
incorporated directly into the searches for gravitational-waves as a parameter
characterizing a candidate event along with the rest of the data from the
gravitational-wave channel, as opposed to a standard approach of vetoing entire
segments of flagged data based on a hard threshold on data quality.

Another advantage of the \ac{MLA} classifiers is that they can incorporate
various, potentially diverse types of information and establish correlations
between multiple parameters.  In addition to the transient noise data used in
this study, we plan to include more slowly-varying baseline information about
the detector subsystems in the future. For example, the quality of alignment in
the interferometer may be important for predicting the amount of noise that
couples into the gravitational-wave channel from elsewhere in the instrument.
Machine learning should be able to automatically identify such non-linear
correlations, even if they are not known previously.

As a final test, we explore several ways of combining the output of several
classifiers in order to increase the robustness of their predictions and
possibly improve combined efficiency. Following general principles for
combing multiple analysis methods, we suggest several approximations for the
optimal combined ranking given by the joint likelihood ratio. We test our
approximations and find that they perform similarly to and do not improve upon
the efficiencies of individual classifiers.

Based on these results, we conclude that the three {MLA} classifiers used in
this study: ANN, SVM, and RF, are all able to achieve robust and competitive
classification performance for our set of data. The RF classifier was the most
robust against the form (range, shape, scaling, number) of input data, while
ANN and SVM benefit from reshaping certain input parameters along physical
arguments. Since all classifiers achieve similar limiting performance and
identify most of the same events, we conclude that they are near-optimal in
their use of the existing data. Future improvement in classification efficiency
is therefore likely to come from including additional sources of useful
information, rather than refinements to the algorithms themselves.

%%%%%%%%%%%%%%%%%%%%%%%%%%%%%%%%%%%%%%%%%%%%%%%%%%%%%%%%%
\acknowledgments
We would like to thank Ryan Fisher for careful reading of the manuscript and many useful comments. We also would like to thank Andrew Lundgren for extensive discussions on the subject of the paper. KK, YMK, CHL, JJO, SHO, and EJS were supported in part by the Global Science experimental Data hub Center (GSDC) at KISTI. KK, YMK, and CHL were supported in part by National Research Foundation Grant funded by the Korean Government (NRF-2011-220-C00029). CHL was supported in part by the BAERI Nuclear R$\&$D program (M20808740002). JC, EOL and XW were supported in part by the Ministry of Science and Technology of China under the National 973 Basic Research Program (grants No.\ 2011CB302505 and No.\ 2011CB302805). TY was supported in part by the National High-Tech Research and Development Plan of China (grant No.\ 2010AA012302). RE, EK and RV were supported by LIGO laboratory. LIGO was constructed by the California Institute of Technology and Massachusetts Institute of Technology with funding from the National Science Foundation and operates under cooperative agreement PHY-0757058
%%%%%%%%%%%%%%%%%%%%%%%%%%%%%%%%%%%%%%%%%%%%%%%%%%%%%%%%

%%%%%%%%%%%%%%%%%%%%%%%%%%%%%%%%%%%%%%%%%%%%%%%%%%%%%%%%%%
\appendix 
%%%%%%%%%%%%%%%%%%%%%%%%%%%%%%%%%%%%%%%%%%%%%%%%%%%%%%%%%%%%%
\section{Derivation of the Decision Surfaces for some MLA Optimization Schemes}%Decision surface derivation for some popular \ac{MLA} optimization criteria}
\label{appendix:fom}
%%%%%%%%%%%%%%%%%%%%%%%%%%%%%%%%%%%%%%%%%%%%%%%%%%%%%%%%%%%%%%

In Section~\ref{detection_problem}, we stress that for the detection (or the two-class classification) problem, the most natural optimizing criterion is the Neyman-Pearson criterion, which requires maximum probability of detection at fixed probability of false alarm. Optimizing this criterion, which in functional form is given by (\ref{Sfunc}), leads to the one-parameter family of decision surfaces defined as surfaces of constant likelihood ratio (\ref{decsurface}). Each decision surface is labeled by a corresponding value of the likelihood-ratio, $\Lambda(\auxvec)$, providing a natural ranking for unclassified events. The higher the likelihood-ratio, the more likely it is that event belongs to Class 1. The likelihood-ratio can be mapped to the particular value of the false alarm probability, $P_{0}(\Lambda)$, which assigns it a statistical significance. In practice, it is often more convenient to define ranking in terms of some monotonic function of the likelihood-ratio, $r(\Lambda)$ (eg: $r(\Lambda) = \ln \Lambda$). Classifying and ranking samples based on the likelihood-ratio is guaranteed to maximize the \ac{ROC} curve $P_{1}(P_{0})$. 

In their standard configurations, most \ac{MLA} classifiers apply other kinds of optimization criteria (e.g. the fraction of correctly classified events or the Gini index). Many of these criteria treat the two classes of the events symmetrically, which often is more appropriate than the asymmetric Neyman-Pearson criterion. In this section, we would like to explore some of the most popular criteria used by the \ac{MLA} classifiers in their relation to the Neyman-Pearson criterion. Specifically, we are interested in establishing consistency between the various criteria used in this study, in that they lead to the same optimal decision surfaces and compatible rankings.      

%%%%%%%%%%%%%%%%%%%%%%%%%%%%%%%%%%%%%%%%%%%%%%%%%%%%%%%%%%%%
\subsection{The Fraction of Correctly Classified Events}
%%%%%%%%%%%%%%%%%%%%%%%%%%%%%%%%%%%%%%%%%%%%%%%%%%%%%%%%%%%
First, we consider probably the most popular criterion - the fraction of correctly classified events. This criterion is used by both \ac{ANN} and \ac{SVM} in our analysis. Following the approach of section~\ref{detection_problem}, we define it as a functional of the decision function, $f(\auxvec)$, on the feature space, $\Vdata$:

\begin{align}
\label{Cdefinition}
C =&    \int_{\Vdata}\! \Theta\left(f(\auxvec) - \Ft \right) p(1 \given \auxvec)p(\auxvec) \diff \auxvec \nonumber \\
 & + \int_{\Vdata}\! \Theta\left(\Ft - f(\auxvec) \right) p(0 \given \auxvec)p(\auxvec) \diff \auxvec \,,
 \end{align}

\noindent where  $p(1 \given \auxvec)$ and $p(0 \given \auxvec)$  are the probabilities for a sample with feature vector $\auxvec$ to be classified as Class 1 and Class 0, respectively, and $p(\auxvec)$ is the probability distribution of obtaining a feature vector, \auxvec, regardless of its class. To elucidate this expression, we note that $p(c\given\auxvec)p(\auxvec)\diff \auxvec$ corresponds to the fraction of total events that fall in the hyper-volume $\diff \auxvec$ and are of class $c$. Without the $\Theta$ functions, these integrals would evaluate to $C = p(1) + p(0) = 1$, and we see that the $\Theta$ functions select only those events that are correctly classified. With this interpretation, $f(\auxvec)$ is the decision function. For a given threshold $\Ft$, it defines two regions of samples as Class 1 and Class 0 through $\Theta\left(f(\auxvec) - \Ft \right)$ and its complement $\Theta\left(\Ft - f(\auxvec) \right)$, respectively. Thus, the first term in (\ref{Cdefinition}) accounts for correctly classified events of Class 1 and the second term does the same for Class 0 events. 

Using Bayes theorem, one can express   $p(1 \given \auxvec)p(\auxvec)$ and $p(0 \given \auxvec)p(\auxvec)$ in the first and second term of (\ref{Cdefinition}) as:

\begin{subequations}
\label{bayesrelations}
\begin{align}
p(1 \given \auxvec)p(\auxvec) = p(\auxvec \given 1)p(1)\,,&\\
p(0 \given \auxvec)p(\auxvec) = p(\auxvec \given 0)p(0) \,.
 \end{align}
\end{subequations}
Here $p(\auxvec \given 1)$ and $p(\auxvec \given 0)$, defined in Section~\ref{detection_problem}, are the probability density functions for feature vectors in the  presence and absence of a glitch in gravitational-wave data, respectively. $p(1)$ and p$(0)$ are the prior probabilities for having a glitch or a clean datum, related to each other via $p(1) + p(0) = 1$ as always. The fraction of correctly classified events is then given by
\begin{align}
\label{Cfinal}
C =& \int_{\Vdata}\! \Theta\left(f(\auxvec) - \Ft \right)p(\auxvec \given 1)p(1) \diff \auxvec \nonumber \\
& +\int_{\Vdata}\!\Theta\left(\Ft - f(\auxvec) \right) p(\auxvec \given 0)p(0) \diff \auxvec \,.
\end{align}
The requirement that the variation of $C$ with respect to $f(\auxvec)$ must vanish,

\begin{align}
\label{Cvariation}
\delta C =& \int_{\Vdata}\! \delta\left(f(\auxvec) - \Ft \right)\delta f(\auxvec) \nonumber \\
            &\times \left[p(\auxvec \given 1)p(1) - p(\auxvec \given 0)p(0)\right] \diff \auxvec = 0 \,.
\end{align}
leads to the following condition for the points, $\auxvect$, satisfying $f(\auxvect) - \Ft = 0$:

\begin{equation}
\label{Ccondition}
p(\auxvect \given 1)p(1) - p(\auxvect \given 0)p(0) = 0\,.
\end{equation}
This equation defines the decision surface consisting of points for which 
\begin{equation}
\label{Csurface}
\Lambda(\auxvect)\frac{p(1)}{p(0)} \equiv \frac{p(\auxvect \given 1)p(1)}{p(\auxvect \given 0)p(0)} = 1\,.
\end{equation} 
Thus, optimizing the fraction of correctly classified events leads to a specific decision surface, for which the likelihood ratio, $\Lambda(\auxvect) = p(0)/p(1)$. By construction, the optimization criterion (\ref{Cdefinition}) treats events of the both classes symmetrically, maximizing the number of correctly classified events. As a result, it selects a specific decision surface for which evidence for the sample to belong to either one class or the other are equal. 

%The prior probabilities $p(1)$ and $p(0)$ are often unknown, and their ratio, $p(1)/p(0)$ could be used as a tuning parameter for the classifier. For example, in our case it can be tuned to set the decision surface at the specific value of the likelihood ratio corresponding to a low value of the probability of false alarm, $P_{0} = 10^{-2}$. In practice, this is done by either re-weighting the training samples or by tuning the configuration parameters of the classifiers (e.g. error margin or cost of misclassified event). The ranks for the samples that are not on the decision surface is defined through the decision function, and they reflect the shape of the decision surface. Although not exact, it is a good approximation for likelihood-ratio rank, especially in the vicinity of the decision surface.

%%%%%%%%%%%%%%%%%%%%%%%%%%%%%%%%%%%%%%%%%%%%%%%%%%%%%%
\subsection{The Gini Index}
%%%%%%%%%%%%%%%%%%%%%%%%%%%%%%%%%%%%%%%%%%%%%%%%%%%%%%

Next, we consider the Gini index criterion, which is used in \ac{RF}. For two-class problems, the Gini index of a region in the feature space, $V$, is defined as 

\begin{equation}
\label{Ginidef}
G(V) = 1 - p^2 -q^2 = 2pq 
\end{equation}
where $p$ and $q$ are fraction of Class 1 and Class 0 samples in the region, with $p+q=1$. The Gini index for multiple regions is given by the average
\begin{equation}
\label{Giniaverage}
G = \sum_{i} G(V_i)p(V_i)
\end{equation}
where $p(V_i)$ is the probability for a sample to be in the region $V_i$. 

For the two-class problem, there are two distinct regions -- the region where samples are classified as Class 1, $V_1$, and the region where samples are classified as Class 0, $V_0$. We make the following definitions:
\begin{subequations}
\label{PQdefinitions}
\begin{align}
P = \int_{\Vdata}\! \Theta(f)p(1\given\auxvec)p(\auxvec) \diff \auxvec = \int_{\Vdata}\! \Theta(f)p(\auxvec\given1)p(1)\diff\auxvec \,,&\\
Q = \int_{\Vdata}\! \Theta(f)p(0\given\auxvec)p(\auxvec) \diff \auxvec = \int_{\Vdata}\! \Theta(f)p(\auxvec\given0)p(0)\diff\auxvec \,,
 \end{align}
\end{subequations}
where P (Q) is the probability that a glitch (clean sample) will fall into $V_1$ and $\Theta(f)$ is shorthand for $\Theta\left(f(\auxvec) - \Ft\right)$. We recognize that the expected fractions of events in $V_1$ can be described as $p=P/\pVg$ and $q=Q/\pVg$, where $\pVg=\int_{\Vdata}\! \Theta(f)p(\auxvec)\diff\auxvec$ is the probability for any event (either glitch or clean sample) to fall in $V_1$. Furthermore, we can immediately write the corresponding relations for $V_0$ in terms of $P$, $Q$, and $\pVg=1-\pVc$. We then obtain:
\begin{equation}
\label{Gfinal}
\frac{G}{2} = \frac{PQ}{\pVg} + \frac{\left(p(1) - P\right)\left(p(0) - Q\right)}{1-\pVg}.
\end{equation}
where $p(1)$ and $p(0)$ are the prior probabilities for an event to be Class 1 and Class 0, respectively. We also note that $\pVg$, $P$, and $Q$ are functionally related through
\begin{subequations}
\label{norm}
\begin{align}
\pVg & \equiv \int_{\Vdata}\! \Theta(f)p(\auxvec) \diff \auxvec \nonumber \\
       & =  \int_{\Vdata}\! \Theta(f)\left[ p(1\given\auxvec)p(\auxvec) + p(0\given\auxvec)p(\auxvec)\right]\diff\auxvec  \nonumber \\ 
       & = P + Q\,.
\end{align}
\end{subequations}

We now optimize the Gini index, (\Cref{Gfinal}), to determine the optimal decision surface. For simplicity, let us first optimize $G$ while holding $p(V_1)$ constant. We are interested in the optimal decision surface's shape, and this optimization will determine it while keeping the ratios of the number of samples in $V_1$ and $V_0$ constant. We obtain
\begin{align}
\label{Gvar}
\frac{1}{2}\frac{\delta G}{\delta f(\auxvec)} & =  \frac{\pt Q + \qt P}{\pVg} + \frac{\pt Q + \qt P}{1-\pVg}\nonumber \\
                                                 & - \frac{ p(1)\qt + p(0)\pt}{1-\pVg}\,,
\end{align}
where
\begin{subequations}
\label{PQvars}
\begin{align}
\pt \equiv \frac{\delta P}{\delta f(\auxvect)} =   \delta_{\auxvec, \auxvect} \int_{\Vdata}\! \delta\left(f(\auxvec) - \Ft\right)p(\auxvec \given 1)p(1) \diff \auxvec \,,&\\
\qt \equiv \frac{\delta Q}{\delta f(\auxvect)} =   \delta_{\auxvec, \auxvect} \int_{\Vdata}\! \delta\left(f(\auxvec) - \Ft\right)p(\auxvec \given 0)p(0) \diff \auxvec \,,
 \end{align}
\end{subequations}
where $\delta_{\auxvec, \auxvect} = \delta f(\auxvec)/\delta f(\auxvect) = \{1\mathrm{\ if\ } \auxvect=\auxvec, 0 \mathrm{\ otherwise}\}$. We see that $\pt$ and $\qt$ are probability density functions defined on the decision surface ($\auxvect \in \{f(\auxvect)=\Ft\}$). We can write these variations in terms of conditional probabilities on the decision surface,
\begin{subequations}
\label{pqt}
\begin{align}
\pt = p(\auxvect \given 1)p(1)  \,,&\\
\qt = p(\auxvect \given 0)p(0) \,.
\end{align}
\end{subequations}

Requiring the variation of the Gini index, (\ref{Gvar}), to vanish leads to the following relation
\begin{equation}
\label{eqforptqt}
\frac{\pt}{\qt} = - \frac{p(1)\pVg - P}{p(0)\pVg - Q}
\end{equation}
Identifying the left hand side of this equation with the likelihood ratio for a point on the decision surface, $\Lambda(\auxvect)(p(1)/p(0))  = p(\auxvect \given 1)p(1)/p(\auxvect \given 0)p(0)$, and using (\ref{norm}) we find that
\begin{equation}
\label{lrforG1}
\Lambda(\auxvect)\frac{p(1)}{p(0)} = 1 \,,
\end{equation}
which will hold for all points on the decision surface. Remarkably, this condition is independent of $P$, $Q$ and $\pVg$. As in the case of the fraction of correctly classified events (\ref{Csurface}), it implies that the optimal decision surface is the surface on which the likelihood ratio is equal to a ratio of the priors.

If we consider the more general maximization problem, in which we allow $\pVg$ to vary, we must maximize
\begin{equation}
\label{Gwithconstraint}
\frac{G}{2} = \frac{PQ}{\pVg} + \frac{\left(p(1) - P\right)\left(p(0) - Q\right)}{1-\pVg} + \lambda \left( \pVg - P - Q \right)\,.
\end{equation}
where we use a lagrange multiplier ($\lambda$) to enforce the condition $\pVg - P - Q = 0$ and consider variations of \pVg to be independent of variations in $f(\auxvec)$. 

The variation with respect to $\pVg$ defines the lagrange multiplier, 
\begin{equation}
\label{lambda}
\lambda = \frac{PQ\left( 1- 2\pVg\right) - \pVg^2\left[ p(1)p(0) - p(1)Q - p(0)P\right]}{\pVg^2{\left(1 - \pVg\right)}^2}
\end{equation}
Variation with respect of $f(\auxvec)$ is given by (\ref{Gvar}) minus $\lambda(\pt+ \qt)$. Setting it to zero for all independent variations of $f(\auxvect)$ leads to a more general condition on the likelihood ratio:
\begin{equation}
\label{lrforG2}
\Lambda(\auxvect) = - \frac{\lambda \pVg\left(1 - \pVg\right) + p(1)\pVg - P}{\lambda \pVg\left(1 - \pVg\right) + p(0)\pVg - Q}\,, 
\end{equation}
which holds separately for all points $\auxvect$ on the decision surface.

First of all, note that this condition still requires a constant likelihood-ratio on the decision surface. For each point on the surface, $\Lambda(\auxvect)$ is determined by $P$ and $Q$, which are constants for a given surface. We recover (\ref{eqforptqt}) and (\ref{lrforG1}) when $\lambda = 0$. In all other cases, the likelihood ratio on the decision surface is given by a quite complicated expression, obtained by plugging (\ref{lambda}) and (\ref{norm}) into (\ref{lrforG2}). In practice, the likelihood ratio on the decision surface is set by the desired value for the probability of false alarm, $P_{0}$, but the ratio will be constant over the entire surface. Optimization of the Gini index, then, will be equivalent to optimizing the \ac{ROC} curve in the region of interest, e.g. in our study near $P_{0} = 10^{-2}$.

\subsection{Asymmetric Criteria}
%%%%%%%%%%%%%%%%%%%%%%%%%%%%%%%%%%%%%%%%%%%%%%%
Both criteria considered so far are symmetric in their treatment of events in Class 1 and Class 0. While the asymmetry can be imposed by tuning the ratio of prior probabilities ($p(1)/p(0)$), in some cases it might be desirable to use an explicitly asymmetric criteria. The \ac{RF} implementation in the StatPatternRecognition package contains two different asymmetric criteria, which we explore in our study: Signal Purity ($P$) and Signal Significance ($S$)~\cite{StatPatternRecognitionPackage}. By construction, they identify one class of events as signal and the other as noise and place more emphasis on correctly classifying signal rather than noise. 

The Signal Purity and the Signal Significance are defined as
\begin{align}
\label{PSdefs}
&P = \frac{\omega_1}{\omega_1 +\omega_0}\,, \\
&S = \frac{\omega_1}{\sqrt{\omega_1^2 +\omega_0^2}}\,,
\end{align}
where $\omega_1$ and $\omega_0$ are the fraction of Class 1 (signal) and Class 0 (noise) events in the signal region. The aim is to identify the signal region with the highest Signal Purity or Signal Significance. In the process of the decision tree construction, the classifier identifies the terminal nodes with the highest values of  $P$ or $S$ and orders nodes using these criteria as a rank. For a given terminal node of a tree, one can express $\omega_1$ and $\omega_0$ in terms of conditional probabilities:
\begin{align}
\label{omegasprobs}
&\omega_1 = p(\auxvec \given 1)\,, \\
&\omega_0 = p(\auxvec \given 0) \,.
\end{align}
In terms of these, one can rewrite the signal purity and the signal significance as
\begin{align}
\label{PSin_probs}
&P = \frac{p(\auxvec \given 1)}{p(\auxvec \given 1) + p(\auxvec \given 0)} = \frac{\Lambda(\auxvec)}{\Lambda(\auxvec) + 1}\,, \\
&S = \frac{p(\auxvec \given 1)}{\sqrt{p(\auxvec \given 1)^2 + p(\auxvec \given 0)^2}} = \frac{\Lambda(\auxvec)}{\sqrt{\Lambda^2(\auxvec) + 1}}\,.
\end{align}
Both quantities are monotonic functions of the likelihood ratio $\Lambda(\auxvec) = p(\auxvec \given 1)/p(\auxvec \given 0)$. Ordering nodes in the tree would be equivalent to ranking events by the likelihood ratio, which in turn is equivalent to using decision surfaces of the constant likelihood ratio for classification. 

In practice, different classifiers have various limitations which results in suboptimal performance. Depending on the application, one algorithm or criteria may be more optimal than another, but we establish here that on a theoretical level they all recover the same optimal solution. In the two-class classification problem, the decision surfaces are surfaces of constant likelihood ratio, which also defines the optimal ranking for samples.

%%%%%%%%%%%%%%%%%%%%%%%%%%%%%%%%%%%%%%%%%%%%%%%%%%%%%%%%%%%%%%%%%%%%%%%%%%
\bibliography{../../bibtex/iulpapers,auxmvcbib}
\end{document}